\documentclass{article}

\usepackage[utf8]{inputenc}
\usepackage{geometry}
\usepackage{graphicx}
\usepackage{bm}
\usepackage{amsmath}
\usepackage[normalem]{ulem}
\usepackage{graphicx}
\usepackage{color,soul}
\usepackage[colorlinks=true, citecolor=black, urlcolor=blue]{hyperref}
\usepackage{natbib}
\usepackage{tabularx}
\usepackage{graphicx}
\usepackage{adjustbox}
\usepackage{geometry}
\usepackage{authblk}
\usepackage{tabularx}
\usepackage[table,xcdraw]{xcolor}
\usepackage{textcomp}
\usepackage{siunitx}
\DeclareSIUnit\atom{atom}
\usepackage{subfig}
\usepackage{tikz}
\usetikzlibrary{shapes}
 \usepackage{multirow}
\usepackage{booktabs}
\usepackage{amssymb}
 \usepackage{amssymb}
 
\begin{document}
\newcommand{\markersquare}{\raisebox{0.5pt}{\tikz{\node[draw,scale=0.4,regular polygon, regular polygon sides=4,fill=none](){};}}}
\newcommand{\markerdiamond}{\raisebox{0.5pt}{\tikz{\node[draw,scale=0.4,diamond,fill=none](){};}}}
\newcommand{\markerround}{\raisebox{0.5pt}{\tikz{\node[draw,scale=0.4,circle,fill=none](){};}}}
\newcommand{\markerone}{\raisebox{0.5pt}{\tikz{\node[draw,scale=0.4,circle,fill=black](){};}}}

\title{Physically Interpretable Interatomic Potentials \textit{via} Symbolic Regression and Reinforcement Learning}

\author[1,2]{Bilvin Varughese}

\author[1,2]{Troy D. Loeffler}
\author[1,2]{Suvo Banik}
\author[1,2]{Aditya Koneru}
\author[1,2]{Sukriti Manna}
\author[2]{Karthik Balasubramanian}
\author[2]{Rohit Batra}
\author[3]{Mathew J. Cherukara}
\author[4]{Orcun Yildiz}
\author[4]{Tom Peterka}
\author[5]{Bobby G. Sumpter}
\author[1,2]{Subramanian K.R.S. Sankaranarayanan\thanks{skrssank@anl.gov, skrssank@uic.edu}}
\affil[1]{Department of Mechanical and Industrial Engineering, University of Illinois, Chicago, Illinois 60607, United States}
\affil[2]{Center for Nanoscale Materials, Argonne National Laboratory, Lemont, Illinois 60439, United States}
\affil[3]{X-ray Sciences Division, Argonne National Laboratory, Lemont, Illinois 60439, United States}
\affil[4]{Mathematics and Computer Science Division, Argonne National Laboratory, Lemont, Illinois 60439, United States}
\affil[5]{Center for Nanophase Material Sciences, Oak Ridge National Laboratory, Oak Ridge, TN 37830, United States}

\maketitle
\begin{abstract}

The development of next-generation molecular simulation models requires moving beyond pre-defined functional forms toward machine learning (ML) techniques that directly capture multiscale physics. Here, we demonstrate such an approach using symbolic regression (SR) with equation learner networks and a reinforcement learning search engine to derive interpretable equations for interatomic interactions. Training data were generated through nested ensemble sampling with density functional theory (DFT) energetics, spanning crystalline to highly disordered states. The optimization of the learner network employed continuous-action Monte Carlo Tree Search (MCTS) combined with gradient descent, enabling efficient exploration of function space. For copper as a representative transition metal, an unconstrained search produced models that outperformed fixed-form Sutton-Chen EAM potentials. The SR-derived models (SR1 and SR2) reproduced key material properties - lattice constants, cohesive energies, equations of state, elastic constants, phonon dispersion, defect formation energies, surface/bulk energetics, and phase transformation with significantly improved accuracy. Furthermore, stringent melting simulations using two-phase solid–amorphous interfaces confirmed that SR models accurately capture the interplay of vibrational entropy, cohesive energy, and structural dynamics, surpassing SC-EAM in both qualitative and quantitative predictions. This highlights the potential of SR to deliver fast, accurate, flexible, and physically meaningful potentials, advancing predictive modeling across scales.

\end{abstract}

\newpage
\section*{Introduction}
Rapid improvements in high-performance computing and algorithmic advances have significantly enhanced our molecular-level understanding of materials and their interfaces\cite{zhang2023graph, zhang2024evolutionary, chakrabarti2021nanoporous}. These developments have paved the way for more accurate and detailed simulations, particularly for low-dimensional systems such as clusters and interfaces, where atomic-scale dynamics play a critical role. However, such systems present unique challenges due to their complex and interconnected dynamical processes, including chemical reactions, atomic and ionic transport, defect chemistry, and solvation dynamics. A fundamental understanding of structure and dynamics in these systems using simulations such as molecular dynamics (MD) relies on an accurate description of atomistic scale interactions, particularly in bulk as well as nanoscale clusters and across interfaces like solid-liquid boundaries\cite{gogoi2019effect, chu2025nonclassical}. 

Substantial progress has been made in empirical force fields (EFFs) tailored to various material classes, but significant limitations remain. EFF models typically express a system's potential energy as the sum of short-range covalent (bonded) and long-range non-bonded interactions. Bonded interactions include bond stretching, angle bending, dihedral torsion, and coordination effects, often categorized into two-body (pairwise), three-body (angle), or four-body (conjugation) terms. However, such classifications vary depending on the research domain (e.g., soft matter versus inorganic systems). Non-bonded interactions encompass electrostatics, dispersion (van der Waals), induction (polarization effects), and exchange (Pauli) repulsion, with many-body contributions extending beyond simple pairwise terms. While existing models such as Stillinger-Weber\cite{stillinger1985computer}, Vashishta\cite{vashishta1990interaction}, Finnis-Sinclair\cite{sutton1990long}, and Tersoff potentials~\cite{backman2012bond, manna2022learning, koneru2022multi, koneru2025development} have achieved success, they often fail to account for dynamic charge transfer and struggle with non-equilibrium systems. For example, the MEAM potential, fitted to equilibrium structures, performs poorly in predicting far-from-equilibrium configurations of Al-Cu alloys\cite{apostol2011interatomic}. To address dynamic charge redistribution during reactive processes, advanced reactive EFFs have emerged, such as the ReaxFF~\cite{keith2010reactive} and Charge-Optimized Many-Body (COMB) potentials\cite{liang2013classical,yu2007charge}. These models integrate bond-order concepts with electronegativity equalization schemes and have gained popularity for their ability to capture complex chemical dynamics. However, these approaches come with their own challenges. Reactive EFFs often require extensive parameterization, involving a large number of independent variables that cannot be directly computed or empirically measured. Parameters are typically fitted using local optimization methods, often framed as least-squares fitting against equilibrium properties such as cohesive energies or elastic constants. Although computationally efficient, such local schemes can become trapped in suboptimal minima, limiting their ability to capture non-stoichiometric defect formation, charge migration, or finite-size effects.

The primary limitation of EFFs lies in their reliance on predefined functional forms to describe interatomic interactions. While these forms have been carefully crafted to capture specific physical and chemical phenomena, they inherently constrain the flexibility of the models. This rigidity makes it challenging to accurately represent complex and diverse chemical environments, particularly in clusters, surfaces, and interfaces, where interactions are highly dynamic and nonlinear. As a result, EFFs struggle to generalize across systems with widely varying atomic configurations and fail to adapt to non-equilibrium processes. In addition to the reliance on fixed functional forms, semi-empirical classical force fields face several other critical shortcomings, including: (1) insufficient training datasets and overemphasis on equilibrium structures, limiting their applicability to non-equilibrium systems; (2) reliance on least-squares fitting, prone to convergence issues and overfitting in high-dimensional parameter spaces; and (3) lack of quantitative cross-validation on independent datasets, reducing their predictive accuracy.
The parameterization of EFF or physics-based interatomic potentials typically involves optimizing their functional forms to replicate material properties obtained from experiments or quantum calculations. The most common approach is local optimization framed as least-squares fitting, where parameters are iteratively adjusted to minimize deviations from target properties such as cohesive energies, elastic constants, or forces. However, this method can encounter issues like convergence difficulties and overfitting, particularly for complex potentials with many parameters. Advanced techniques such as genetic algorithms and Bayesian optimization are employed to address these challenges. Genetic algorithms\cite{koza1992programming,brown2010efficient,slepoy2007searching,kenoufi2015symbolic,makarov1998fitting} use an evolutionary approach to explore the parameter space, avoiding local minima, while Bayesian optimization\cite{liu2008bayesian,dequidt2015bayesian,mcdonagh2019utilizing} applies probabilistic modeling to balance exploration and refinement efficiently. Similarly, Monte Carlo sampling\cite{iype2013parameterization,cosseddu2017force} provides a means to explore vast parameter spaces randomly, although it can be computationally intensive.

To enhance accuracy and prevent overfitting, regularization techniques introduce penalties for overly complex parameter sets\cite{cawley2007preventing,wagner2016theory}, and multi-objective optimization is used to balance competing goals, such as matching both energies and forces\cite{koneru2022multi, koneru2024ab,koneru2023multi, koneru2024machine, chan2019machine,zhang2021multi}. Validation, often through cross-validation on independent datasets, ensures the generalizability of the parameterized potentials beyond the training data. Recent advancements incorporate machine learning (ML) to guide parameterization, where ML models predict error landscapes or optimize parameter selection. These approaches, particularly in conjunction with traditional methods, have allowed for improved efficiency and robustness in handling the complexities of interatomic interactions.

Machine learning (ML)-based interatomic potentials\cite{zuo2020performance,nyshadham2019machine,artrith2017efficient,shapeev2016moment,huan2017universal}, such as neural networks\cite{behler2007generalized,varughese2024active}, Gaussian Approximation Potentials (GAP)\cite{bartok2010gaussian,banik2024development}, and Spectral Neighbor Analysis Potentials (SNAP)\cite{thompson2015spectral}, provide a viable alternative to the EFFs by offering much higher accuracy and flexibility. We note that the EFF models based on fundamental physical relationships are significantly simpler and faster than machine-learning (ML) potentials, enabling simulations over longer time and length scales\cite{zuo2020performance}. Their physics-based nature allows them to perform reasonably well in unfamiliar local environments, even those outside their training data. While they require less data to train due to a smaller parameter space, the simplicity of their pre-defined functional form often limits their accuracy compared with the higher precision achievable with ML-based potentials. On the other hand, the ML models often face challenges related to relatively higher computational cost and interpretability. ML models in many cases act primarily as interpolation tools, and while modern MLIPs often include uncertainty quantification to flag out-of-distribution predictions, their accuracy ultimately depends on the diversity of the training dataset; a limited dataset increases the risk of overfitting and reduced transferability to unseen configurations. Another critical issue is their limited interpretability—unlike physics-based models, ML potentials do not inherently provide insights into the underlying physical interactions, making it difficult to derive mechanistic trends or validate predictions. Moreover, the absence of embedded physical constraints in neural network-based potentials can result in unphysical predictions, especially in extreme non-equilibrium conditions that are not included in the training dataset.
Recent developments in machine learning interatomic potentials (MLIPs) have introduced formally complete frameworks such as Moment Tensor Potentials (MTP)\cite{shapeev2016moment} and the Atomic Cluster Expansion (ACE)\cite{drautz2019atomic,ortner2023atomic} These approaches provide systematic and mathematically complete basis expansions that, in principle, can span the full functional space of atomic interactions and encompass many existing MLIP formalisms. However, while MTP and ACE offer completeness and extensibility, the interpretability of their large sets of expansion coefficients remains limited, often requiring nontrivial post hoc analysis to extract insights such as the relative weight of low- versus high-order body terms. 

Symbolic regression (SR) presents a promising alternative to the ML models\cite{wang2024exploring,schmidt2009distilling,udrescu2020ai}, allowing the discovery of functional forms directly from data without relying on predefined equations. Unlike traditional ML approaches, SR represents learned mappings as physically meaningful equations, enabling the extraction of chemical trends and correlations\cite{guo2022improving,zhang2023optimizing,tan2022discovery}. SR has been successfully applied to develop phenomenological models and discover interatomic potentials that accurately describe interactions in molecular dynamics simulations. For example, a type of SR is genetic programming (GP)\cite{kenoufi2015symbolic}, which is an evolutionary algorithm that optimizes mathematical expressions by simulating natural selection. It has been successfully applied to learn interatomic potentials by deriving symbolic representations of the potential energy surface (PES)\cite{kenoufi2015symbolic,hernandez2019fast,hernandez2023generalizability} directly from ab initio data. GP starts with a population of random equations and iteratively refines them through selection, crossover, and mutation based on their ability to reproduce target energies and forces. This method enables the discovery of fast, accurate, and transferable many-body potentials without relying on predefined functional forms. By producing interpretable and physically meaningful equations, GP bridges the gap between empirical models and more complex machine-learning-based approaches, offering a powerful tool for modeling atomic interactions.

In this work, we adopt symbolic AI to learn physical equations directly from first-principles datasets, bypassing the limitations of predefined functional forms. Using symbolic regression, specifically equation-learning networks, we perform an unconstrained exploration of function spaces defined by mathematical operators, analytical functions, material properties, and constants. An equation learner network is a specialized framework that combines neural network capabilities with symbolic representations to discover interpretable mathematical relationships in data. These networks integrate neural network architectures with symbolic representations to search for equations, optimizing both their structure and parameters to derive accurate and physically meaningful models, such as those for interatomic potentials. Our approach enables us to derive functional forms for molecular dynamics simulations – here, we focus on a representative transition metal like copper and demonstrate the significant improvements in performance achieved by the SR learnt physics model. Our methodology aims to bridge the gap between traditional empirical force fields and ML models, offering both interpretability and predictive accuracy in describing atomistic interactions.

\section*{Results}

\subsection*{Benchmarking of EqNN - Recovery of the Sutton Chen Model}

We first benchmark and validate our algorithm by learning the functional form of a known potential energy surface (PES) as defined by the Sutton Chen Embedded Atom Model (SC-EAM) for copper. An active learning approach is adopted, where the training starts with a small set of structures near equilibrium. Next, both the energy and force predictions are refined iteratively using a nested ensemble sampling approach (see Methods). Using a relatively sparse dataset of near-ground state and far-from-ground state structures (total $~$ 341), we trained the equation neural network (EqNN) to discover an SR model that closely approximates the Sutton Chen energy model. The training dataset was progressively enhanced by identifying structures that exhibited prediction errors beyond a defined threshold \textbf{(110\%} of current training error) and adding them to the training set. This active learning process, repeated over many iterations, allowed for a systematic refinement of both the dataset and the predicted equation, until the reference PES was accurately reproduced.

\begin{figure*}[ht]
\centering

\includegraphics[width =1.0\textwidth]{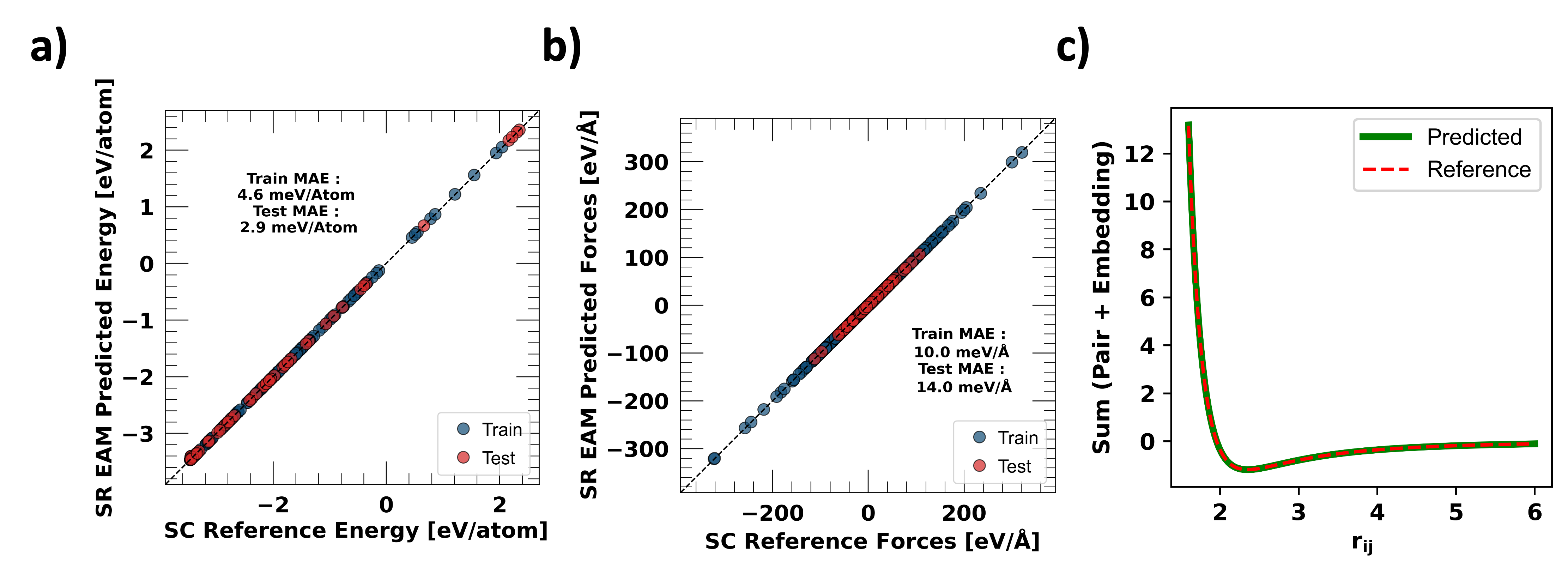}
\caption{ The figure captures the training process for learning the Sutton-Chen Equation using EqNN a) The energy and force correlations showing very high correlations between the reference energy and force value calculated using SC EAM vs the ones predicted using EqNN b) The comparison between the pair term and the embedding term for the target equation and the one predicted by EqNN where for all relevant distances. c) The comparison of the Reference vs the predicted equations using the EqNN
}
\label{fig:training_process}
\end{figure*}

 The Monte Carlo Tree Search (MCTS) framework played a critical role in exploring the functional space and guiding the refinement of our EqNN model. The comparison between the Sutton-Chen reference equation and the final predicted equation (Fig.~\ref{fig:training_process}c) demonstrates how closely the learned equation approximates the original. The pair, density, and embedding terms, initially described by the Sutton Chen model as:

\begin{equation}
\begin{aligned}
E_{\text{pair}}(r_{ij}) &= \frac{1289.3050}{r_{ij}^{9}},\\[1mm]
\rho(r_{ij}) &= \frac{527.6214}{r_{ij}^{6}},\\[1mm]
F_{\text{emb}}(\rho) &= -1.0\,\sqrt{\rho}.
\end{aligned}
\end{equation}

were successfully predicted by our EqNN as:

\begin{equation}
\begin{aligned}
E_{\text{pair}}(r_{ij}) &= \frac{1292.5470}{r_{ij}^{9}},\\[1mm]
\rho(r_{ij}) &= \frac{859.0663}{r_{ij}^{6}},\\[1mm]
F_{\text{emb}}(\rho) &= -0.7852\,\sqrt{\rho}.
\end{aligned}
\end{equation}

Fig.~\ref{fig:training_process}a) provides a quantitative assessment of our SR model's performance by plotting predicted vs. reference (SC-EAM) energies and forces. The SR model achieved a mean absolute error (MAE) of \SI{4.6}{\milli\electronvolt\per\atom} for training energy and \SI{2.9}{\milli\electronvolt\per\atom} for test energy. The parity plot showing force predictions (Fig.~\ref{fig:training_process}b) showed an MAE of \SI{10.0}{\milli\electronvolt\per\angstrom}\ for training and \SI{14.0}{\milli\electronvolt\per\angstrom}\ for testing, demonstrating high accuracy in force reproduction as well. These results highlight the robustness of the EqNN in capturing both energy and force behavior of the Sutton Chen EAM potential. Finally, in Fig.~\ref{fig:training_process}c, we compare the sum of predicted embedding term and pair term against the reference SC-EAM equations over relevant pairwise distances. The near-perfect overlap between the two models for all distances shows that the EqNN was able to accurately learn and reproduce the Sutton Chen model’s behavior, indicating the success of our training approach in capturing the essential physics of the target PES.

\subsection*{Limitations of the SC-EAM Model}

The SC-EAM model is a widely used potential for modeling the properties of transition metals, particularly copper (Cu). SC-EAM is popular due to its ability to accurately capture the essential physics of metallic bonding, including cohesive energy, equilibrium structure, and elastic properties. It balances computational efficiency with reliable performance, making it a common choice for large-scale simulations of Cu and other face-centered cubic (FCC) metals. The SC-EAM model is particularly effective for studying bulk properties, surfaces, and defect dynamics in Cu. Additionally, its computational simplicity has contributed to its adoption in simulations of nanostructures, thin films, and interfaces involving Cu. Despite these strengths, the model has limitations in describing highly directional bonding or extreme non-equilibrium conditions, where there is significantly high configurational diversity.

\begin{figure*}[ht]
\centering
\includegraphics[width =1.0\textwidth]{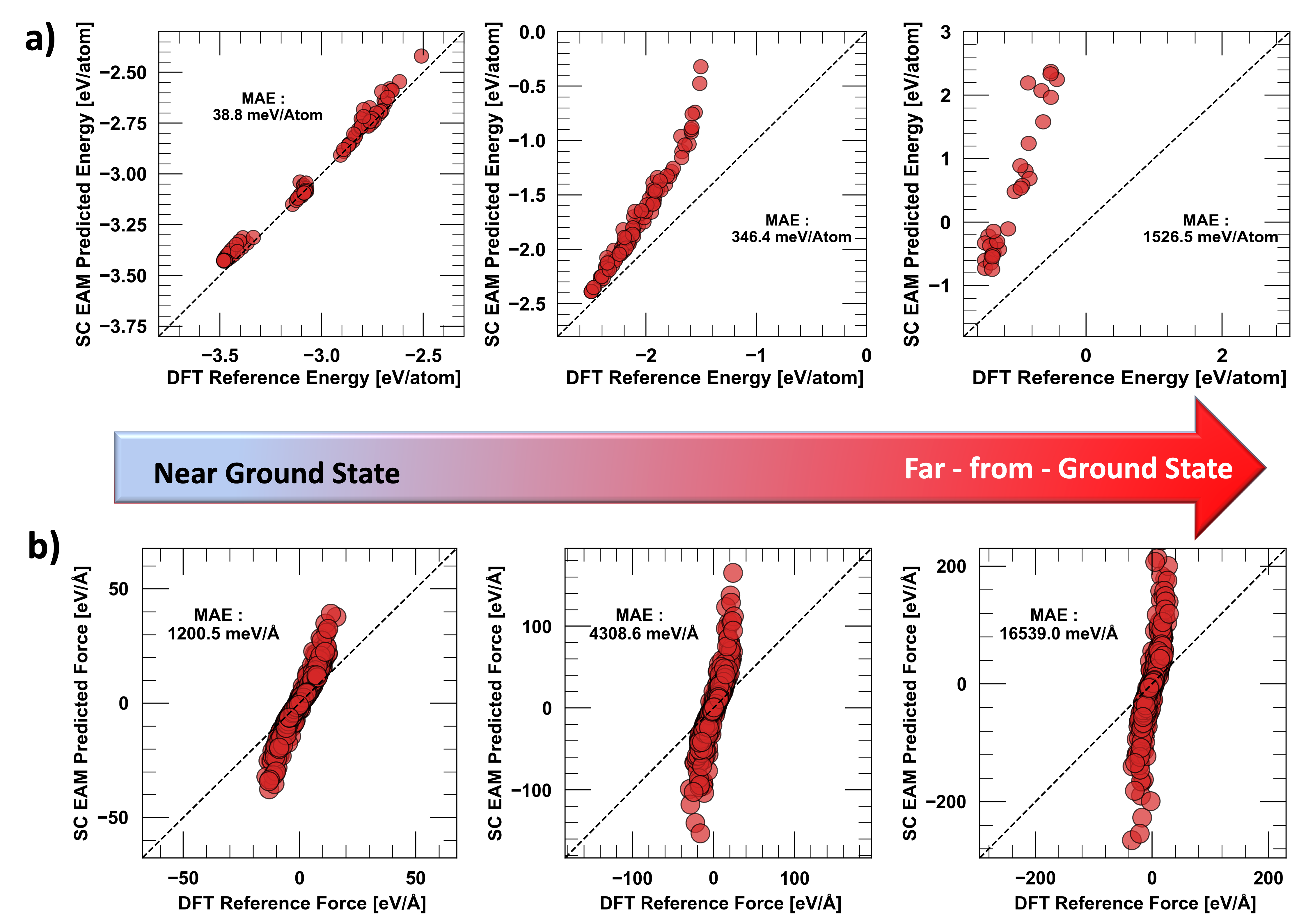}
\caption{Comparison of SC EAM predictions with DFT reference data for structures used to train the EqNN corresponding to the SC Equation. (a) Energy and (b) force predictions are shown across configurations ranging from near-ground state (left) to far-from-ground state (right). As the configurations deviate from equilibrium, both energy and force prediction errors (MAE) increase significantly.
 }
\label{fig:SCvalidation_results}
\end{figure*}

First, we performed a comparative analysis between the energy and force predictions of the SC-EAM model and the corresponding DFT calculations on the current dataset. This comparison enabled us to assess the performance of the SC-EAM across a spectrum of configurations, including both equilibrium and non-equilibrium states.
Fig.~\ref {fig:SCvalidation_results} represents a parity plot showing a comparison between the predictions of the SC-EAM and reference DFT values for the dataset used to train the equation neural network (EqNN). The plots reveal an increase in both energy and force prediction errors as we move away from equilibrium configurations. For structures near equilibrium (leftmost panel), the SC-EAM performs well, with a mean absolute error (MAE) of \SI{38.8}{\milli\electronvolt\per\atom} for energy and \SI{1200.5}{\milli\electronvolt\per\angstrom}\ for forces. However, as we progress towards higher-energy configurations (rightmost panels), the predictive performance of SC-EAM significantly deteriorates, with MAE values reaching \SI{1526.5}{\milli\electronvolt\per\atom} for energy and \SI{16539.0}{\milli\electronvolt\per\angstrom}\ for forces.

These results underscore the limitations of traditional potential models like SC-EAM, particularly in capturing system behavior in non-equilibrium states. The observed degradation in predictive accuracy suggests the need for more sophisticated models capable of accurately describing a broader range of configurations and the need for the development of more generalizable interatomic potentials that are crucial for accurately simulating materials in diverse conditions. Building on the success of our algorithm in identifying the functional forms of many-body potentials, we next turn our focus towards evaluating its capability to infer potential models from density functional theory (DFT)-derived data.

\subsection*{Development of SR model with a fixed Embedding using EqNN and DFT training dataset }

\paragraph{SR1 Model Development}

The inability of the SC-EAM model to describe the highly non-equilibrium configurations stems from the lack of flexibility in its functional form. First, we seek to improve the model flexibility by allowing the pair and density terms to be discovered through our EqNN algorithm, while fixing the embedding function to \( \sqrt{\rho} \). A key feature of our partially constrained search approach thus involved fixing the embedding term, commonly used in the SC-EAM formalism, while allowing the remaining terms to evolve. By anchoring the embedding function to \( \sqrt{\rho} \), we retained its known effectiveness in capturing many-body interactions while enhancing the flexibility of the pair and density functions. Using Monte Carlo Tree Search (MCTS) guided by mean absolute error (MAE) as the scoring function, we explored the functional space for the pair and density terms. The tree search of the EqNN weights was solely informed by data generated from density functional theory (DFT), providing a robust foundation for learning a more flexible and accurate interatomic potential for copper. The functional forms learned by the EqNN algorithm reflect complex terms that go beyond the limitations of traditional pair potentials, particularly when modeling non-equilibrium configurations. The learned potential forms were as follows:

\begin{equation}
\begin{aligned}
E_{\text{pair}}(r_{ij}) &= 0.0390\,\exp\Biggl[
-\frac{0.0088}{r_{ij}^{10}}
-\frac{2.4339}{r_{ij}^{9}}
+\frac{10.1231}{r_{ij}}\\[1mm]
&\quad\quad
-\frac{0.2701}{r_{ij}^{12}}
-\frac{3.1511}{r_{ij}^{6}}
-0.0490\,\exp\bigl(1.5418\,r_{ij}\bigr)
\Biggr],\\[2mm]
\rho(r_{ij}) &= \frac{8.9043}{r_{ij}^{7}}
+\frac{132.5196}{r_{ij}^{6}}
+\frac{0.6619}{r_{ij}^{5}}
+2.5025\,\exp\Bigl(-1.3632\,r_{ij}\Bigr),\\[2mm]
F_{\text{emb}}(\rho) &= -1.5598\,\sqrt{\rho}.
\end{aligned}
\end{equation}

\begin{figure}[ht]
\centering
\includegraphics[width =1.0\textwidth]{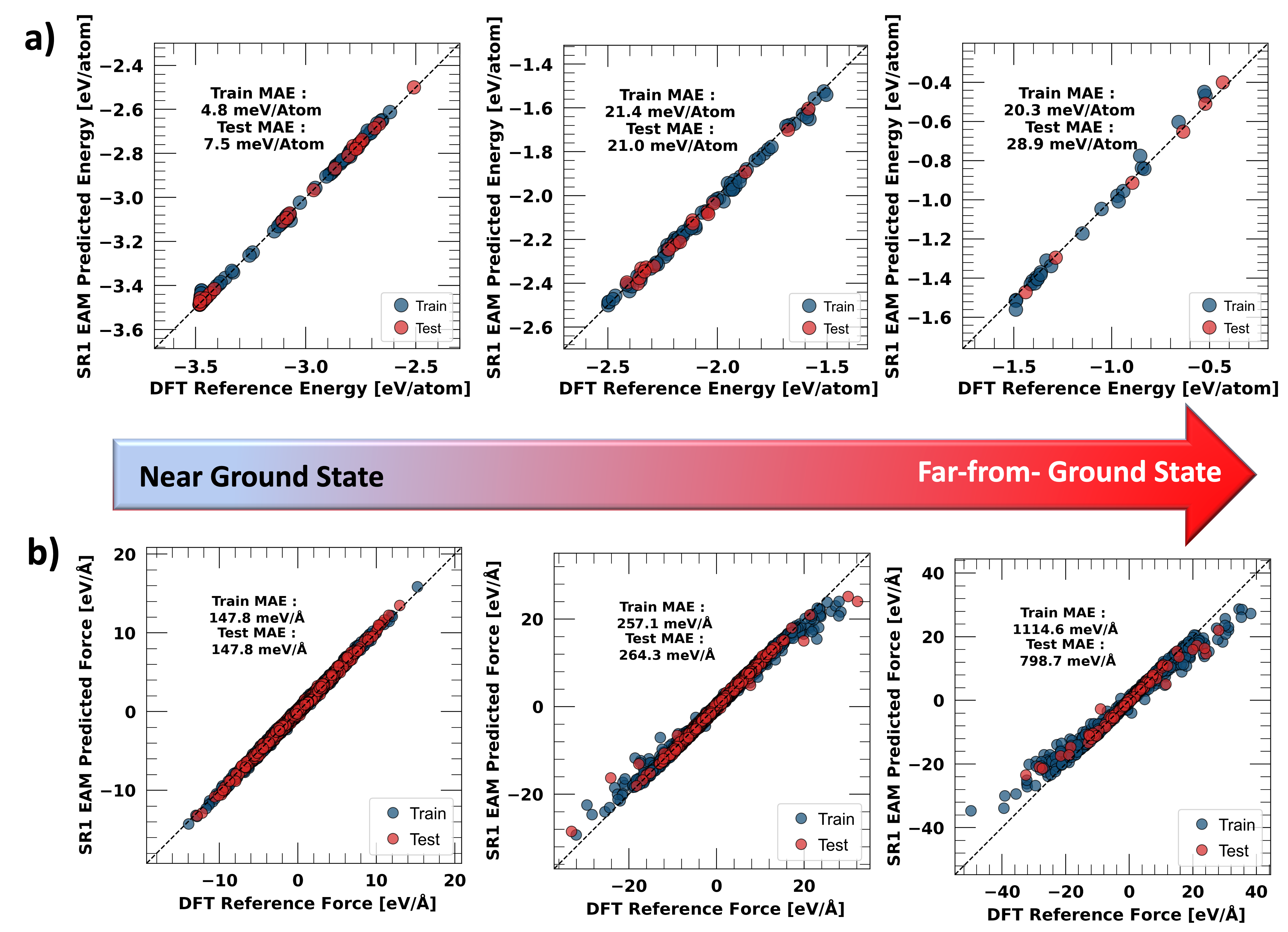}
\caption{Comparison of SR1 EAM predictions with DFT reference data for the dataset used to train the EqNN, using the same fixed embedding as the SC-EAM formalism. (a) Energy and (b) force predictions are shown across configurations ranging from near-ground state (left) to far-from-ground state (right). SR1 EAM exhibits significantly improved correlation with DFT energies and forces across all regimes compared to SC-EAM.}

\label{fig:SR1}
\end{figure}

\paragraph{Energies and Force Comparison}

To validate the model, we further compare the predictions made by the new SR1 EAM potential against the reference DFT data. As seen in Fig.~\ref{fig:SR1}, the SR1 EAM demonstrates a significant improvement in both energy and force predictions compared to the SC-EAM model. For near equilibrium structures, the MAE for energy predictions was reduced to \SI{4.8}{\milli\electronvolt\per\atom} for the SR1 EAM, a substantial improvement over the \SI{38.8}{\milli\electronvolt\per\atom} error of the SC-EAM. This trend holds for non-equilibrium structures as well, with the SR1 EAM maintaining significantly lower error margins across all energy ranges.

A similar pattern is observed in the force predictions. Near-ground state structures under SC-EAM yielded force prediction errors as high as \SI{1200}{\milli\electronvolt\per\angstrom}, whereas the SR1 EAM reduced this error to \SI{147.8}{\milli\electronvolt\per\angstrom} for similar structures. Even for more complex, non-equilibrium configurations, the SR1 EAM consistently outperforms the SC-EAM, with error reductions visible across the entire dataset. The improvements in force predictions are particularly important for accurately capturing dynamical interactions, especially in materials modeling applications where forces drive structural evolution and dynamical transitions. One of the key insights from this comparison is the ability of the SR1 EAM to generalize well across different structural regimes. Whereas the SC-EAM struggled to maintain accuracy as structures deviated from equilibrium, the SR1 EAM, leveraging the learned functional forms for the pair and density terms, accurately predicted the energy and force for a wide variety of configurations. This robustness suggests that the combination of fixed embedding with flexible functional terms offers a powerful strategy for improving interatomic potentials.

\paragraph{Equation of State, Phonon Dispersion, and Elastic Constants}
\begin{figure}[ht]
\centering
\includegraphics[width =0.80\textwidth]{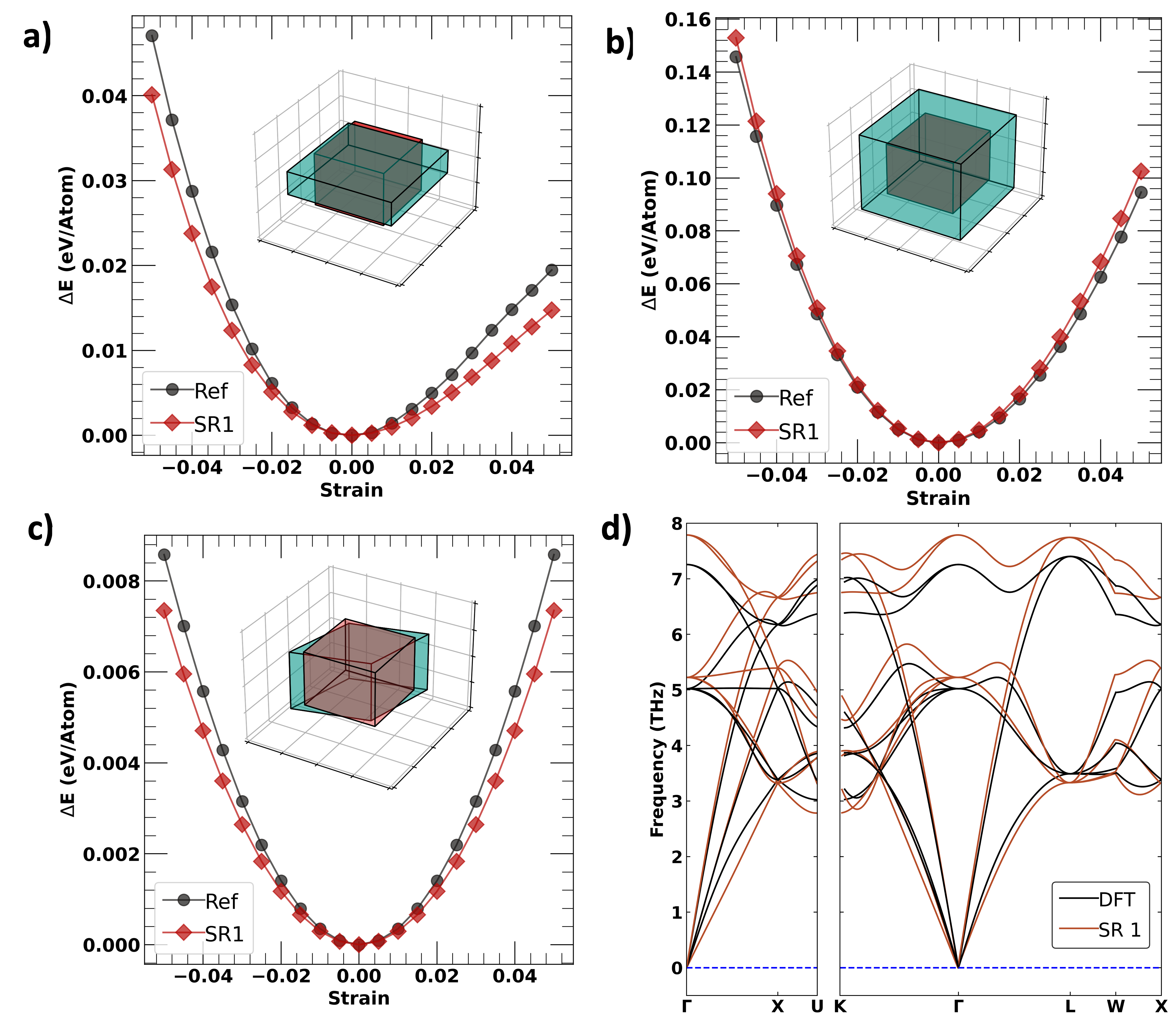}
\caption{ (a) Uniaxial equation of state (EOS) under compressive/tensile strain: SR1 (red) closely follows reference DFT (Ref; gray/black) across the full strain window, with a mean absolute error (MAE) of \(5.26\,\mathrm{meV/atom}\).
(b) Volumetric EOS: SR1 maintains strong correlation with DFT (MAE \(=2.89\,\mathrm{meV/atom}\)).
(c) Shear/tetragonal deformation: SR1 again agrees well with DFT (MAE \(=5.4\,\mathrm{meV/atom}\)).Insets in (a–c) depict the applied deformation modes.
(d) Phonon dispersion along \(\Gamma\!-\!X\!-\!U\!-\!K\!-\!\Gamma\!-\!L\!-\!W\!-\!X\): SR1 reproduces the DFT branches with good agreement, particularly the acoustic slopes near \(\Gamma\); no imaginary modes are observed and a slight under-prediction of frequencies is noted.
All energies are reported as \(\Delta E\) per atom referenced to unstrained fcc.
}
\label{fig:SR1EOS}
\end{figure}

We further validate the SR1 model by calculating its equation of state (EOS) and comparing it with that obtained using DFT. To compute the EOS, which describes the energy change with deformation, we subject the system to an energy minimization. The simulation box is then systematically deformed by scaling its volume isotropically (normal deformation) or anisotropically (shear deformation). For isotropic deformations, the simulation box is uniformly scaled across all dimensions to simulate hydrostatic compression or expansion. In contrast, anisotropic deformations can involve altering the dimensions along one or two axes, which simulate uniaxial or biaxial strain. Shear deformations are applied by distorting the box shape (e.g., altering angles between axes), which mimics shear stresses. At each deformation step, the atomic positions are relaxed while keeping the deformation fixed, and the total potential energy is computed. By systematically repeating this process across a range of deformations, we plot the energy as a function of deformation. Shear and normal deformation analyses help in understanding the elastic or plastic behavior of the material.  

Fig.~\ref{fig:SR1EOS} highlights the performance of the SR1 model in accurately predicting the EOS across all tested deformation modes, including both normal and shear stresses, as benchmarked against reference DFT calculations. In panel~(a), the material's response under uniaxial compressive and tensile stresses is presented. The SR1 potential demonstrates excellent agreement with DFT across the full range of strain, achieving MAE of \SI{5.26}{\milli\electronvolt\per\angstrom}. Panel~(b) illustrates the material's behavior under volumetric stresses, where SR1 maintains a high degree of correlation, with an overall MAE of \SI{2.89}{\milli\electronvolt\per\angstrom}. These EOS evaluations significantly contribute to the predictions of elastic constants \(C_{11}\) and \(C_{12}\), with SR1 achieving predictive errors of 5 \% and 10 \%, respectively, with respect to the experimental values.

Panel~(c) focuses on the material's response to shear stress. Here, the SR1 potential exhibits strong agreement with DFT, with an average error of 5.4~meV/atom. However, due to the high sensitivity of \(C_{44}\) to the EOS under shear deformation, the SR1 model incurs a larger error of 20.9 \% in \(C_{44}\) with respect to experiments. Despite this, the performance of SR1 represents a significant improvement over the SC-EAM potential, underscoring SR1's robustness in modeling complex deformation modes.

We compare the elastic constants for the SC-EAM and SR1 models relative to those obtained using DFT. Table \ref{tab:bulk_mechanical_properties} also shows the experimental values for the various elastic constants. Note that the SC-EAM model was primarily fit to bulk equilibrium properties and does quite well in predicting the elastic constants ($C_{11}$, $C_{12}$, and $C_{44}$) when compared to reference values. The SR1 model improves the performance in predicting  $C_{44}$
but shows a slight increase in $C_{11}$ and $C_{12}$ error (170 GPa and 130 GPa for SC vs. 166.94 GPa and 112.27 GPa for SR1). Nonetheless, the predicted elastic constants overall show very good agreement with DFT as well as experimental values.

Another key metric for evaluating the predictive accuracy of a force field, particularly for crystalline solids, is the comparison of phonon dispersion relations with ab initio calculations. Phonon dispersion relations represent the dependence of the normal mode frequencies, $\omega$ (\textbf{k}, j), on the reciprocal space wavevector \textbf{k} for all branches j and specific crystallographic directions. Phonon dispersions were obtained via the finite-displacement method using Phonopy \cite{parlinski1997first,phonopy-phono3py-JPCM,phonopy-phono3py-JPSJ}. DFT reference forces were computed with VASP, and potential-based forces were evaluated via LAMMPS. Similar to the predictions of the SC-EAM, the SR1 also displays reasonable accuracy, and the acoustic branches match well with those of reference DFT - a slight under-prediction is observed, suggesting that they are weaker than those calculated by DFT. Nonetheless, the predictions and slope of the various acoustic branches at the \textbf{$\Gamma$} point are in good agreement with DFT.

\subsection*{Development of a fully flexible SR model using EqNN and DFT training dataset}

Building on the improved performance seen above, we explored the EqNN for further improvement by extending the search space to include new embedding functions. By allowing the algorithm to be more flexible and discover more optimal embedding functions through functional space exploration, we aim to refine the model’s accuracy beyond the current level, particularly in predicting behaviors under non-equilibrium conditions.

\paragraph{Model Development}
\begin{figure}[ht]
\centering
\includegraphics[width =1.0\textwidth]{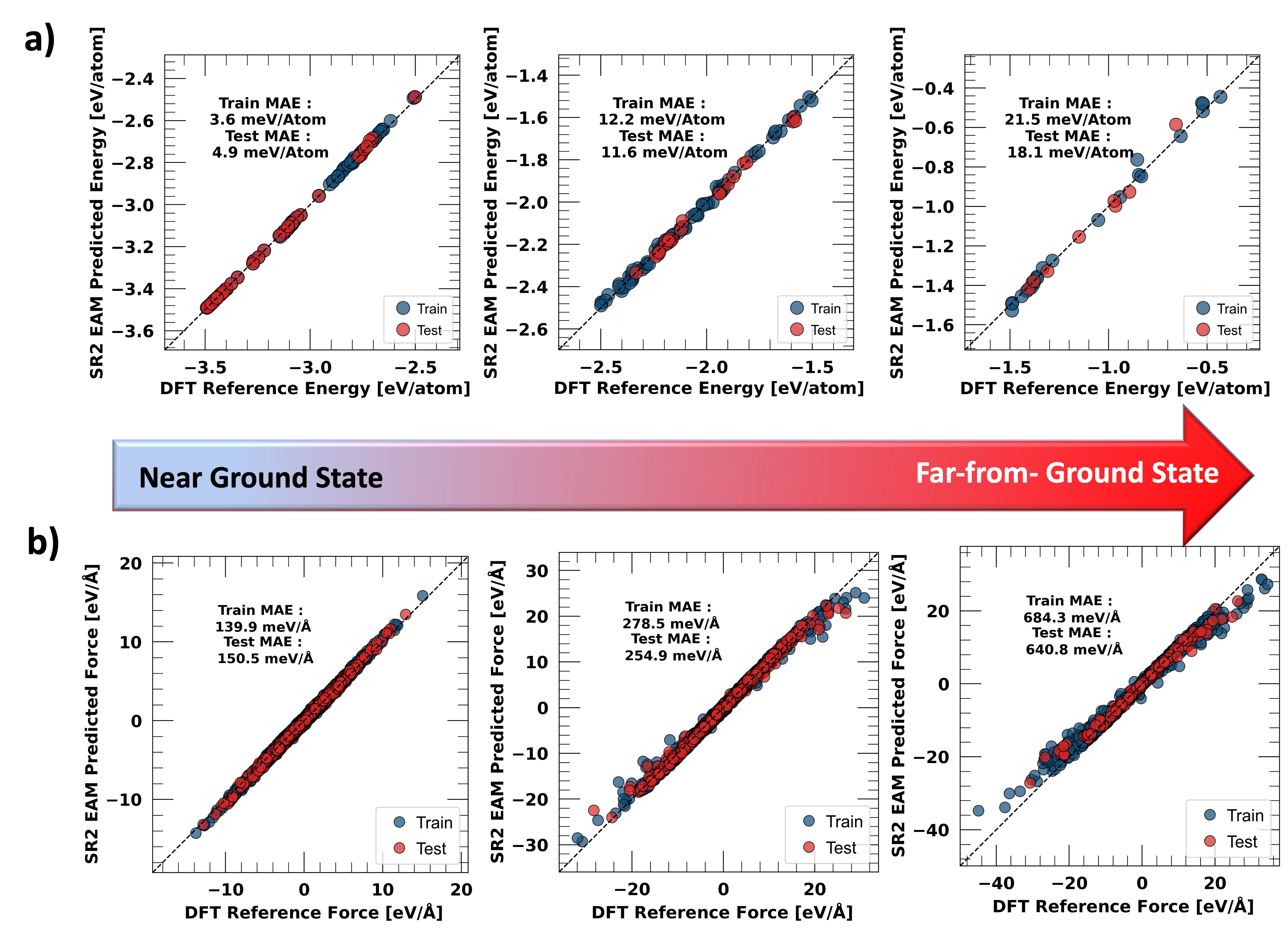}
\caption{Comparison of SR2 EAM predictions with DFT reference data for the dataset used to train the EqNN, using the embedding function having more terms in addition to the square root term in the basis set. (a) Energy and (b) force predictions are shown across configurations ranging from near-ground state (left) to far-from-ground state (right). From left to right, we can see an increase in the range of energies where these structures fall, and we see a higher correlation in both energy and forces for all energy ranges compared with the predictions made by SR1. }
\label{fig:SR2vsDFT}
\end{figure}
We introduced additional flexibility to the embedding function, allowing the Monte Carlo Tree Search (MCTS) algorithm to explore a broader functional space as shown in Table \ref{tab:basis_functions}. Following the search protocol outlined in the Methods section, we search over an expanded set of basis functions for the embedding term to further improve upon the performance in the near as well as far-from-equilibrium regions.

We converged to a preliminary network of a loss of $\sim$ \SI{80}{\milli\electronvolt\per\atom} after 55000 MCTS steps, following which another 10000 steps using Adam optimizer were performed, giving us the final SR2 model. We observe that the optimized embedding term now includes a quadratic and linear correction term in \( \rho \), alongside the original \( \sqrt{\rho} \), improving the model’s accuracy. The new equation for the SR2 EAM potential is:

\begin{equation}
\begin{aligned}
E_{\text{pair}}(r_{ij}) &= 0.0396\,\exp\Biggl[
-\frac{0.0088}{r_{ij}^{10}}
-\frac{2.4353}{r_{ij}^{9}}
+\frac{10.1196}{r_{ij}}\\[1mm]
&\quad\quad
-\frac{0.2730}{r_{ij}^{12}}
-\frac{3.1447}{r_{ij}^{6}}
-0.0490\,\exp\bigl(1.5427\,r_{ij}\bigr)
\Biggr],\\[2mm]
\rho(r_{ij}) &= \frac{9.0593}{r_{ij}^{7}}
+\frac{133.8980}{r_{ij}^{6}}
+\frac{0.6663}{r_{ij}^{5}}
+2.5137\,\exp\Bigl(-1.3516\,r_{ij}\Bigr),\\[2mm]
F_{\text{emb}}(\rho) &= -1.5788\,\sqrt{\rho}
-0.0009\,\rho^{2}
+0.0175\,\rho.
\end{aligned}
\end{equation}

We note that the newly discovered SR2 model further significantly enhances both energy and force predictions, particularly for structures far from equilibrium. When compared to the SR1 and SC-EAM models, the SR2 EAM shows a marked reduction in errors, especially in highly non-equilibrium structures. The parity plots showing the comparison of the SR2 predicted energies and forces with respect to those from DFT are shown in Fig. \ref{fig:SR2vsDFT}. We observe that the MAE for force predictions drops substantially, with improvements visible across the board, particularly for the most complex configurations. For instance, the force error for non-equilibrium structures decreased from \SI{1114.6}{\milli\electronvolt\per\angstrom} in SR1 to \SI{684.3}{\milli\electronvolt\per\angstrom} in SR2, highlighting the SR2 model's capability to better capture the underlying physical interactions. This enhanced accuracy in force predictions is critical for modeling atomic-scale interactions in systems under stress or deformation, where accurate forces dictate the evolution of the structure. The reduction of error in non-equilibrium configurations is especially significant for practical applications such as defect formation, diffusion, and phase transitions, where accurate forces and energies are paramount for reliable simulations. The improved predictive power of SR2 EAM thus marks a critical advancement in the development of more flexible and generalizable interatomic potentials.

\paragraph{Equation of State, Phonon Dispersion and Elastic Constants}
\begin{figure}[ht]
\centering
\includegraphics[width =0.8\textwidth]{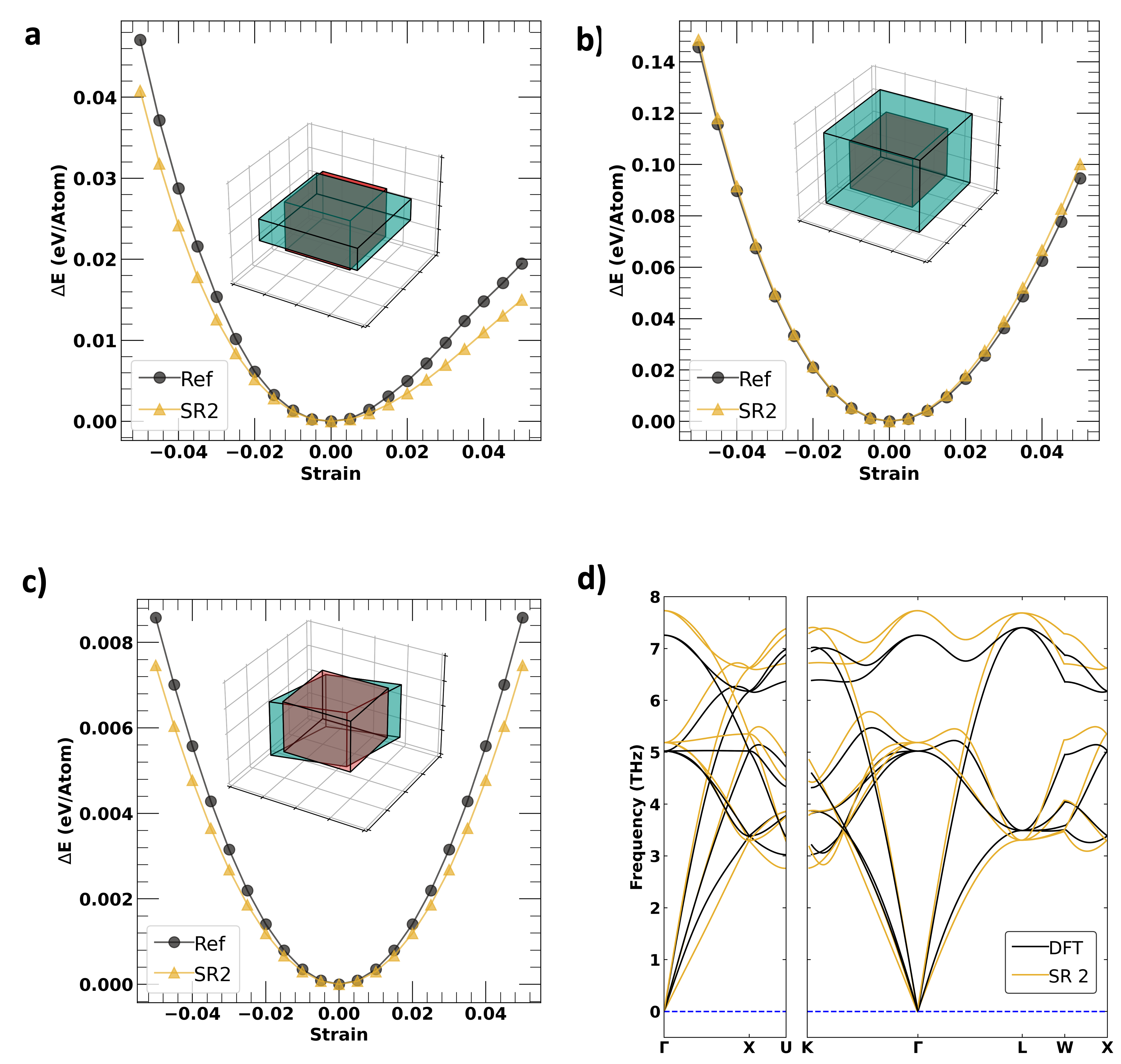}
\caption{(a–c) Change in energy per atom (\(\Delta E\)) versus strain for three deformation modes—(a) uniaxial, (b) volumetric (isotropic), and (c) shear showing SR2 (gold) closely tracking reference DFT (Ref; black) across \(\pm\,5 \%\) strain. Insets in (a–c) depict the applied deformation modes. (d) Phonon dispersion along \(\Gamma\!-\!X\!-\!U\!-\!K\!-\!\Gamma\!-\!L\!-\!W\!-\!X\): SR2 reproduces DFT with excellent agreement on the acoustic slopes near \(\Gamma\) and no imaginary modes, and overall shows improved correspondence relative to SR1. All energies are referenced to the unstrained fcc state. }
\label{fig:SR2EOS}
\end{figure}
We further assessed the performance of the fully flexible SR2 model by calculating its equation of state (EOS) under various deformations. Fig.~\ref{fig:SR2EOS} shows that the model accurately predicts the EOS across all deformation modes. For both the isotropic (normal) and anisotropic (shear) deformation, we observe that the SR2 model performs on par with that of the SR1 and SC-EAM models when compared to DFT. Additionally, calculations of the elastic constants shown in Table \ref{tab:bulk_mechanical_properties} revealed low errors, affirming the SR2 model’s effectiveness in capturing copper's mechanical properties. In particular, the predicted elastic constants have 3 $\%$ error in $C_{11}$, 6.2 $\%$ error in $C_{12}$, and 16 $\%$ error in $C_{44}$ when compared with DFT. The $C_{11}$ and $C_{12}$ performance is comparable to that of SR1 and SC-EAM. We note a significant improvement in the $C_{44}$ when compared to SC-EAM, which has an error of 27 $\%$ when compared to DFT. Overall, the performance of the SR2 and SR1 models in predicting elastic constants represents an improvement over the SC-EAM model. Finally, we also compare the phonon dispersion predicted by the SR2 model with respect to DFT. We find that the SR2 model displays high accuracy, and the acoustic branches are in excellent agreement with those of reference DFT. Overall, the phonon dispersion predictions for the SR2 model show an improvement over the SR1 model owing to its improved flexibility.

\begin{figure*}[ht]
\centering
\includegraphics[width =0.70\textwidth]{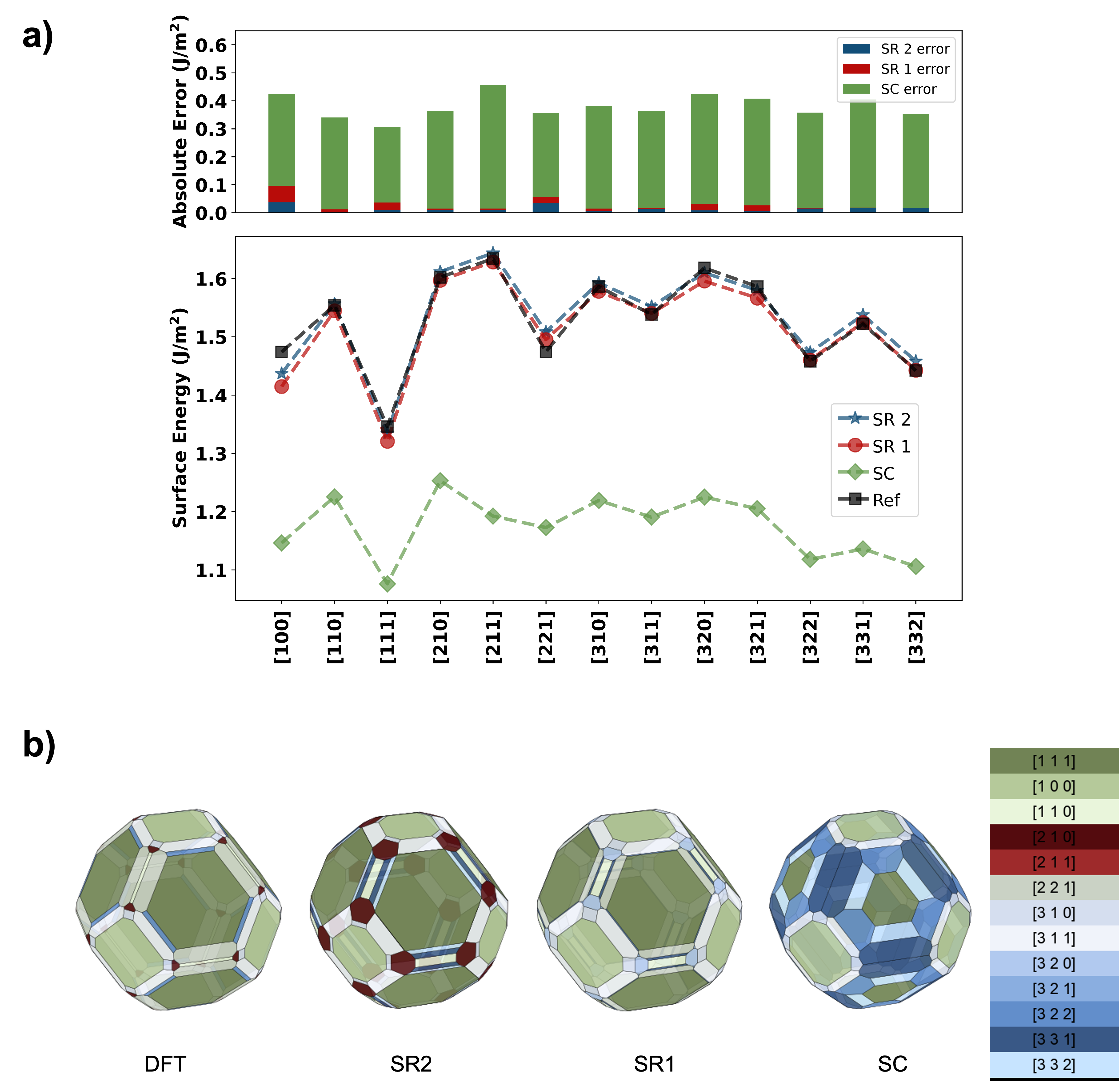}
\caption{The surface energy predictions. (a) Orientation-resolved surface energies \(\gamma(hkl)\) (bottom) for DFT (Ref) and the three force fields (SR2, SR1, SC), together with the corresponding absolute errors \(|\gamma_{\text{model}}-\gamma_{\text{DFT}}|\) (top). Across the set of low- and higher-index facets shown on the abscissa, SR2 and SR1 closely track the DFT values, while SC exhibits systematically larger deviations. Energies are in \(\mathrm{J\,m^{-2}}\).
(b) Equilibrium Wulff constructions derived from the same \(\gamma(hkl)\) datasets. SR2 and SR1 reproduce the DFT polyhedron with similar facet types and relative areas, whereas SC shows noticeable changes in facet weighting. Facet colors correspond to the \([hkl]\) families indicated in the legend.}
\label{fig:SurfaceEnergyPredictions}
\end{figure*}
\begin{table}[ht]
\centering
\caption{Performance of Symbolic Network potentials on predicting bulk mechanical properties.}
\begin{tabular}{lccccc}
\toprule
Property & DFT & SC & SR1 & SR2 & Expt \\ 
\midrule
$C_{11}$ (GPa) & 180 & 170 & 166.94 & 186.21 & 176\cite{simmons1971handbook} \\ 
$C_{12}$ (GPa) & 127 & 130 & 112.27 & 119.11 & 125\cite{simmons1971handbook} \\ 
$C_{44}$ (GPa) & 79  & 58  & 64.84  & 66.40  & 82\cite{simmons1971handbook} \\ 
$SFE$ ($mJ/m^2$) & 41\cite{branicio2013effect}  & 32.45 & 49.84  & 66.40  & \textemdash \\ 
$E_{1v}$ (eV)  &     & 0.875 & 1.32 & 1.37 & 1.29\cite{triftshauser1975monovacancy} \\
$E_{1i}$ (eV)  & 2.902\cite{Chaturvedi2022} & 2.72 & 3.08 & 3.02 & \textemdash \\ 
\bottomrule
\end{tabular}
\label{tab:bulk_mechanical_properties}
\end{table}

\subsection*{Performance Assessment - Surface Energetics and Ordering}
Accurate surface energy predictions enable reliable simulations of phenomena such as crystal growth, fracture mechanics, adhesion, wetting, and catalytic activity, to name a few. An interatomic potential that accurately models surface energies ensures that computational studies can faithfully reproduce and predict material behaviors involving surfaces, leading to a better understanding and design of materials interfaces. We therefore assess the performance of the SR1, SR2, and the SC-EAM model in predicting surface energetics across a variety of low and high-index surfaces. The Cu(111) surface is the most stable and often studied for equilibrium properties and stability, whereas the other low-index surfaces (100) and (110) are relevant for adsorption, reconstruction, and catalytic studies. The (311), (210), (310), and other planes with higher Miller indices are less stable and have more under-coordinated atoms, but are relevant in catalytic applications and designing nanostructures. Note that the SR1 and SR2 potentials included surface structures in their training to achieve precise predictions of surface energies. In comparison with the SC-EAM model, our models show notably lower errors relative to reference DFT computed surface energies and capture the trends of surface energies across different orientations with high fidelity. In particular, both the SR1 and SR2 models show excellent quantitative agreement for the low-index surfaces and capture the trends across the entire range of low and high-index planes. This is illustrated in Fig.~\ref{fig:SurfaceEnergyPredictions}, where panel (a) demonstrates the close quantitative agreement between surface energy predictions from our EqNN-derived potentials (SR1 and SR2) and DFT, significantly outperforming the SC-EAM in accuracy as well as the trends observed between the various low and high-index surfaces.

Capturing Wulff constructions is essential in materials applications as they provide a fundamental understanding of a crystal's equilibrium shape based on its surface energies. The Wulff construction predicts the most energetically favorable morphology a crystal will adopt by relating the surface energy of each crystallographic facet to its distance from the crystal's center. This geometric representation allows us to anticipate crystal growth and the shapes it will form under equilibrium conditions, which is crucial for applications where crystal shape affects performance. Accurately modeling Wulff constructions is particularly important for designing materials at the nanoscale, where surface effects significantly influence properties like catalytic activity, optical behavior, and mechanical strength. By understanding and predicting how different facets contribute to the overall crystal shape, one can tailor materials for specific applications, such as enhancing catalytic efficiency or controlling nanostructure assembly in advanced technologies. We thus evaluate the performance of the various models in predicting the Wulff construction. Panel (b) presents Wulff constructions to visualize equilibrium crystal shapes based on surface energies predicted by each model. We include all available surface indices in the analysis. The SR2 model shows a closer agreement with DFT compared to even SR1 - the increased flexibility in SR2 allows it to capture surface energetics slightly better than SR1. Overall, we find that both SR1 and SR2 maintain consistency with DFT predictions even across a broad range of surface orientations and represent a significant improvement over the SC-EAM model. These results highlight the ability of our SR1 and SR2 models to not only predict accurate surface energies but also effectively capture realistic surface energy trends as well as Wulff constructions, which are critical for reliable morphology and stability predictions in materials research.

\subsection*{Performance Assessment - Bulk Mechanical Properties, Grain Boundary Energetics, Bain Transformation Path, and Thermal Expansion Coefficients}

Our symbolic-regression potentials (SR1, SR2) accurately reproduce both planar-fault and point-defect energetics in Cu with small, systematic errors. From Table~\ref{tab:bulk_mechanical_properties}, the intrinsic stacking-fault energy $\gamma_{\mathrm{SF}}$ is $43.5$ and $43.3~\mathrm{mJ\,m^{-2}}$ for SR1 and SR2, deviating from the DFT reference ($41~\mathrm{mJ\,m^{-2}}$) by $+6.1\%$ and $+5.6\%$, respectively. For point defects, the vacancy formation energy $E_{1v}$ from SR1/SR2 (1.32/1.37~eV) closely matches experiment (1.29~eV), with modest errors of $+2.3\%$ and $+6.2\%$, whereas the SC EAM prediction (0.875~eV) substantially underestimates $E_{1v}$ ($-32\%$). Interstitial formation energies $E_{1i}$ of 3.08/3.02~eV (SR1/SR2) lie closer to the DFT benchmark (2.91~eV), with relative deviations of $+5.8\%$ and $+3.8\%$, and improve upon the Sutton--Chen (SC) model (2.72~eV; $-6.5\%$ vs.\ DFT). Collectively, SR1/SR2 recover stacking-fault energetics within $\sim\!6\%$ of DFT and deliver quantitatively reliable vacancy and interstitial formation energies—surpassing SC and aligning with first-principles and experimental benchmarks.
\begin{table*}[h]
\centering
\small
\caption{Cu grain boundaries with rotation axis, angle, plane, DFT GB formation energy $\gamma_{\rm GB}^{\rm DFT}$ and predictions from SR1, SR2, and SC-EAM (J\,m$^{-2}$).}
\begin{tabular}{c c c c c r r r r}
\toprule
$\Sigma$ & Type & Rotation axis & Rotation angle ($^\circ$) & GB plane & $\gamma_{\rm GB}^{\rm DFT}$ & $\gamma_{\rm GB}^{\rm SR1}$ & $\gamma_{\rm GB}^{\rm SR2}$ & $\gamma_{\rm GB}^{\rm SC-EAM}$ \\
\midrule
3  & twist & $[111]$ & 60.00  & $(111)$ & 0.07 & 0.002 & 0.002 & 0.000 \\
7  & twist & $[111]$ & 38.21  & $(111)$ & 0.37 & 0.351 & 0.355 & 0.221 \\
11 & tilt  & $[110]$ & 50.48  & $(113)$ & 0.37 & 0.352 & 0.356 & 0.218 \\
57 & tilt  & $[110]$ & 97.05  & $(445)$ & 0.42 & 0.350 & 0.354 & 0.239 \\
41 & tilt  & $[110]$ & 55.88  & $(443)$ & 0.45 & 0.450 & 0.454 & 0.335 \\
3  & tilt  & $[110]$ & 109.47 & $(112)$ & 0.63 & 0.687 & 0.695 & 0.435 \\
33 & tilt  & $[110]$ & 58.99  & $(225)$ & 0.65 & 0.676 & 0.683 & 0.416 \\
27 & tilt  & $[110]$ & 148.41 & $(115)$ & 0.72 & 0.776 & 0.783 & 0.509 \\
51 & tilt  & $[110]$ & 16.10  & $(551)$ & 0.74 & 0.795 & 0.803 & 0.511 \\
5  & twist & $[100]$ & 36.87  & $(100)$ & 0.75 & 0.794 & 0.803 & 0.479 \\
17 & tilt  & $[110]$ & 86.63  & $(223)$ & 0.78 & 0.796 & 0.806 & 0.533 \\
43 & tilt  & $[110]$ & 80.63  & $(335)$ & 0.84 & 0.884 & 0.894 & 0.561 \\
3  & tilt  & $[111]$ & 180.00 & $(110)$ & 0.85 & 2.981 & 3.026 & 2.671 \\
9  & tilt  & $[110]$ & 38.94  & $(221)$ & 0.86 & 0.957 & 0.968 & 0.570 \\
5  & tilt  & $[100]$ & 53.13  & $(013)$ & 0.88 & 0.946 & 0.957 & 0.593 \\
7  & tilt  & $[111]$ & 38.21  & $(321)$ & 0.92 & 0.988 & 1.000 & 0.633 \\
9  & tilt  & $[110]$ & 38.94  & $(114)$ & 0.92 & 0.939 & 0.948 & 0.609 \\
27 & tilt  & $[110]$ & 148.41 & $(552)$ & 0.93 & 1.006 & 1.017 & 0.619 \\
19 & tilt  & $[110]$ & 26.53  & $(331)$ & 0.94 & 0.947 & 0.958 & 0.605 \\
17 & tilt  & $[110]$ & 93.37  & $(334)$ & 0.97 & 0.979 & 0.990 & 0.604 \\
19 & tilt  & $[110]$ & 26.53  & $(116)$ & 0.98 & 1.057 & 1.069 & 0.674 \\
5  & tilt  & $[100]$ & 36.87  & $(021)$ & 1.00 & 1.066 & 1.079 & 0.692 \\
66 & tilt  & $[110]$ & 20.05  & $(118)$ & 1.03 & 1.079 & 1.091 & 0.681 \\
9  & twist & $[110]$ & 38.94  & $(110)$ & 1.17 & 1.242 & 1.257 & 0.789 \\
\bottomrule
\end{tabular}
\label{tab:grain Boundary_properties}
\end{table*}
Table \ref{tab:grain Boundary_properties} further benchmarks grain-boundary (GB) formation energies in Cu across diverse geometries (varying $\Sigma$ values, tilt/twist characters, rotation axes, and GB planes). The SR models reproduce DFT $\gamma^{\mathrm{GB}}$ with high fidelity, typically within $5$--$10\%$ relative error: for example, at $\Sigma 11$ ($\gamma^{\mathrm{DFT}}_{\mathrm{GB}}=0.37~\mathrm{J\,m^{-2}}$), SR1/SR2 predict $0.352/0.356~\mathrm{J\,m^{-2}}$ (errors $<5\%$), whereas SC underestimates more strongly ($0.218~\mathrm{J\,m^{-2}}$, $-41\%$); at $\Sigma 17$ tilt ($\gamma^{\mathrm{DFT}}_{\mathrm{GB}}=0.78~\mathrm{J\,m^{-2}}$), SR1/SR2 give $0.796/0.806~\mathrm{J\,m^{-2}}$ (errors $<4\%$) compared with SC at $0.533~\mathrm{J\,m^{-2}}$ ($-32\%$). One notable outlier is the high-angle $\Sigma 3$ tilt case, where SR1/SR2 overshoot DFT ($\gamma^{\mathrm{DFT}}_{\mathrm{GB}}=0.85~\mathrm{J\,m^{-2}}$) with values near $3.0~\mathrm{J\,m^{-2}}$. Aside from this exception, SR1 and SR2 exhibit near-DFT accuracy across complex interfacial environments, often deviating by less than $5$--$10\%$, whereas SC errors can exceed $30$--$40\%$, underscoring the superior transferability of the symbolic-regression potentials.

\begin{figure}[h]
  \centering
  \includegraphics[width=0.9\linewidth]{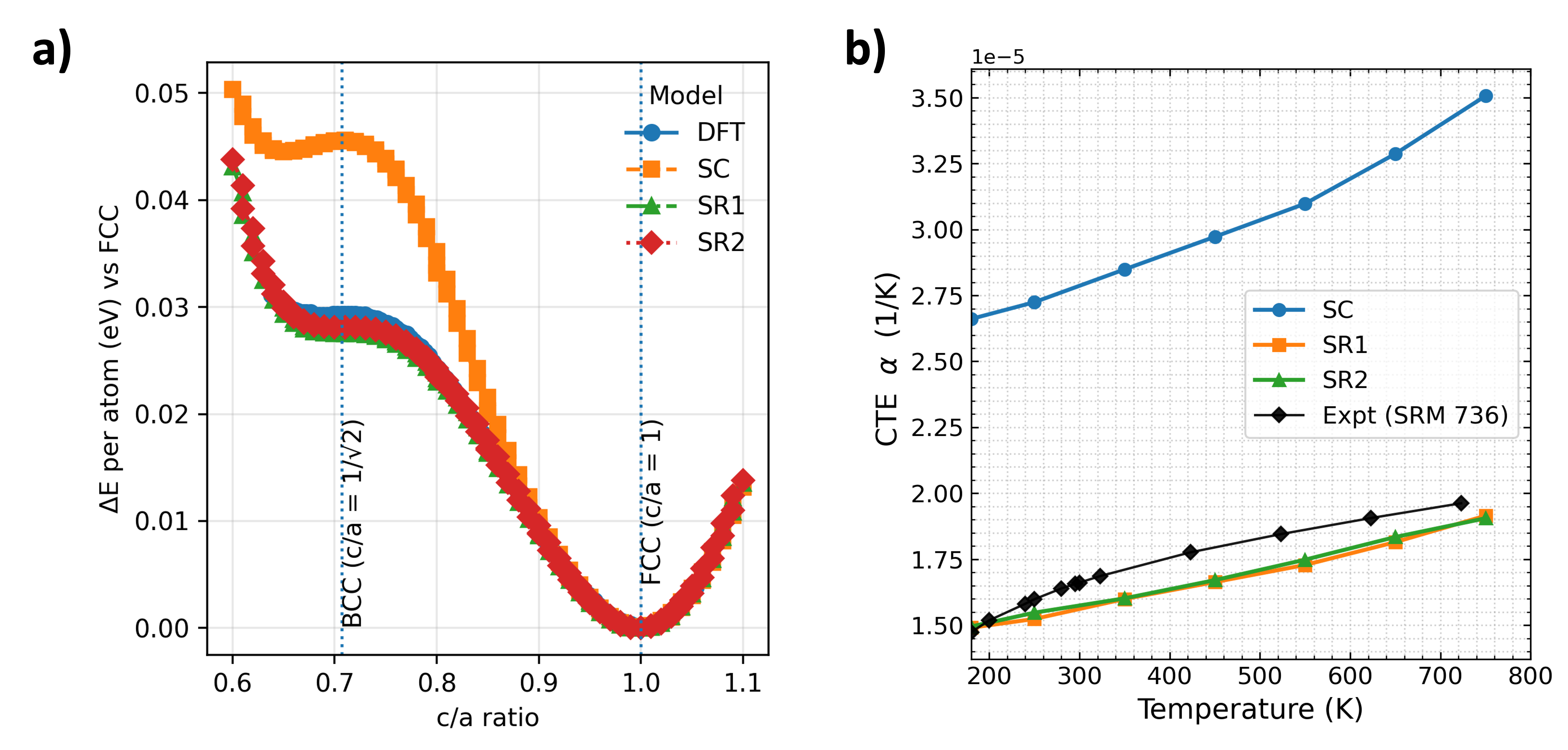}
  \caption{(a) Bain transformation path in Cu: energy per atom relative to fcc vs.\ $c/a$. All models give a minimum at fcc ($c/a=1$) and a maximum at bcc ($c/a=1/\sqrt{2}$); SR1/SR2 follow DFT closely, while SC overestimates the barrier. (b) Linear CTE, $\alpha \times 10^{6}$ (K$^{-1}$), vs.\ temperature: SR1/SR2 agree with SRM~736 within $\sim$5--10\% from 200–750~K, whereas SC is $\sim$40--70\% higher; curves are truncated at 750~K for comparability. }

  \label{fig:cte-Bein}
\end{figure}

We also evaluate the Bain transformation path for copper—a homogeneous tetragonal deformation that continuously connects fcc and bcc by varying the axial ratio \(c/a\) at (approximately) fixed volume. In this parameterization, \(c/a=1\) corresponds to fcc, while \(c/a=1/\sqrt{2}\) yields bcc via a body-centered tetragonal (bct) configuration. The resulting energy profile probes both the relative phase stability and the transformation barrier. Fig.~\ref{fig:cte-Bein} (a) plots the energy per atom (relative to fcc) versus \(c/a\). Density-functional theory (DFT) establishes fcc (\(c/a=1\)) as the global minimum and bcc (\(c/a=1/\sqrt{2}\)) as a metastable maximum. At the FCC reference point, SR1 and SR2 coincide with the DFT energy to numerical accuracy (0\% error), whereas the Sutton--Chen (SC) potential exhibits an artificial offset of \(\sim 5\)~meV/atom. At the bcc maximum, DFT predicts \(\sim 0.029\)~eV/atom; SR1 and SR2 reproduce this barrier with only \(+3\%\) and \(+7\%\) deviations, respectively, while SC overestimates it at \(\sim 0.045\)~eV/atom (\(+55\%\)). Thus, the SR models faithfully capture both the qualitative features and the quantitative energetics of the Bain path, in sharp contrast to the large systematic errors of SC.

We further evaluate the potentials’ performance on thermal properties by computing copper’s linear coefficient of thermal expansion (CTE) from room temperature to $800$~K. The CTE quantifies the fractional change in a characteristic length $L$ with temperature $T$ at constant pressure, $\alpha = (1/L)\,(\partial L/\partial T)_P$, and provides a sensitive test of a potential’s thermomechanical transferability. Fig.~\ref{fig:cte-Bein} reports the temperature dependence of Cu’s CTE up to $800$~K. The Sutton--Chen (SC) potential substantially overestimates $\alpha$ across the range, yielding values that are $\sim$40--70\% higher than the SRM~736 experimental reference. In contrast, the symbolic-regression models (SR1, SR2) closely track experiment, remaining typically within 5--10\% of SRM~736 between $200$ and $750$~K. Beyond matching the magnitude, SR1/SR2 also reproduces the observed temperature trend, whereas SC exhibits an unrealistically steep increase. These results indicate markedly improved thermomechanical fidelity and transferability for SR1/SR2 relative to the empirical SC baseline.

\subsection*{Performance Assessment - Equations of state for polymorphs}

\begin{figure}[h]
  \centering
  \includegraphics[width=0.9\linewidth]{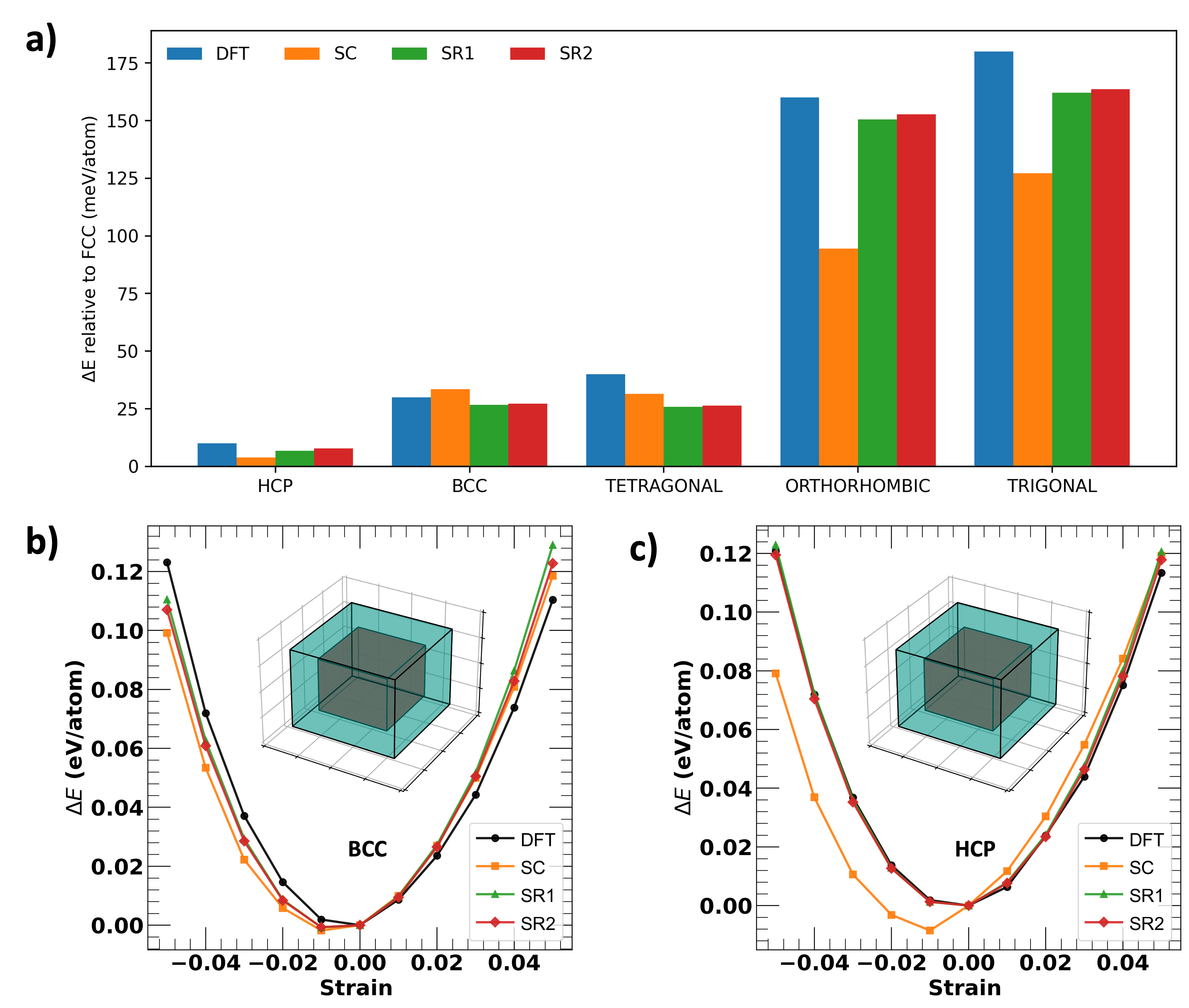}
  \caption{ Comparison of model performance on polymorph properties. a) Polymorph energies relative to fcc: SR1/SR2 reproduce DFT ordering and magnitudes more accurately than SC, especially for higher-energy orthorhombic and trigonal phases. (b–c) Equation-of-state curves \((\Delta E\ \text{vs.}\ \text{strain})\) for BCC and HCP (eV/atom). 
SR1 and SR2 closely follow DFT near the minima and across the sampled strain range, while SC shows larger deviations, particularly for HCP. Insets illustrate the applied deformation.}
  \label{fig:BCC-HCP EOS}
\end{figure}

We also extended our validation to include multiple polymorphs beyond FCC (HCP, BCC, Tetragonal, Orthorhombic, Trigonal, etc.). As shown in Fig.~\ref{fig:BCC-HCP EOS}(a), the SR potentials correctly reproduce the expected energetic ordering with all phases higher in energy than FCC. The predicted energy differences are within the error limits and follow the correct trends, showing improved accuracy compared to the standard Sutton–Chen EAM.
We also compare the Volumetric EOS for Phases other than FCC, like BCC and HCP, for all three potentials vs DFT. And we see that for both BCC and HCP, the potentials developed using symbolic regression perform excellently vs DFT. The equilibrium strain (minimum of the energy--strain curve) is located at $\sim$0 for DFT, SR1, and SR2, while the SC potential shows a small offset ($\sim$0.5--1\% volumetric strain), reflecting a mismatch in equilibrium volume.
At the energy minimum, SC predicts a value shifted upward by 3--4~meV/atom relative to DFT (an error of $\sim$15--20\%), while SR1 and SR2 reproduce the reference within $\sim$1~meV/atom ($<$5\% error).
Furthermore, the curvature of the SC curve is underestimated compared to DFT, indicating an underprediction of the bulk modulus.
By contrast, SR1 and SR2 accurately capture both the equilibrium energy and the curvature across the full strain range ($\pm$5\%), demonstrating near-DFT fidelity in describing volumetric deformation for BCC and HCP phases.

\subsection*{Performance Assessment - Prediction of Melting Dynamics}

Prediction of thermodynamic properties such as melting point is critical for assessing the reliability of any force field used in MD simulations because it directly reflects the force field's ability to capture the fundamental interatomic interactions that govern phase transitions. Accurate prediction of the melting point ensures that the force field correctly models the thermodynamic stability of a material and its response to temperature changes, both of which are crucial for studying processes such as solidification, nucleation, crystal growth, and phase stability. Melting involves a complex interplay of vibrational entropy, cohesive energy, and structural dynamics, making it a stringent test of a force field's accuracy. A force field capable of accurately predicting the melting point demonstrates that it can represent both the anharmonic interactions and the energy landscape transitions necessary for modeling high-temperature behavior. Additionally, the melting point is an excellent benchmark for validating and tuning force fields to ensure they are not over-parameterized to specific properties but generalizable across thermodynamic conditions. In practical terms, materials simulations often involve conditions near the melting point, such as sintering, high-temperature deformation, and heat treatment. A force field that fails to predict the melting point accurately may yield unreliable results in these scenarios, limiting its applicability. 

\begin{figure*}[p]
\centering
\includegraphics[width =1.0\textwidth]{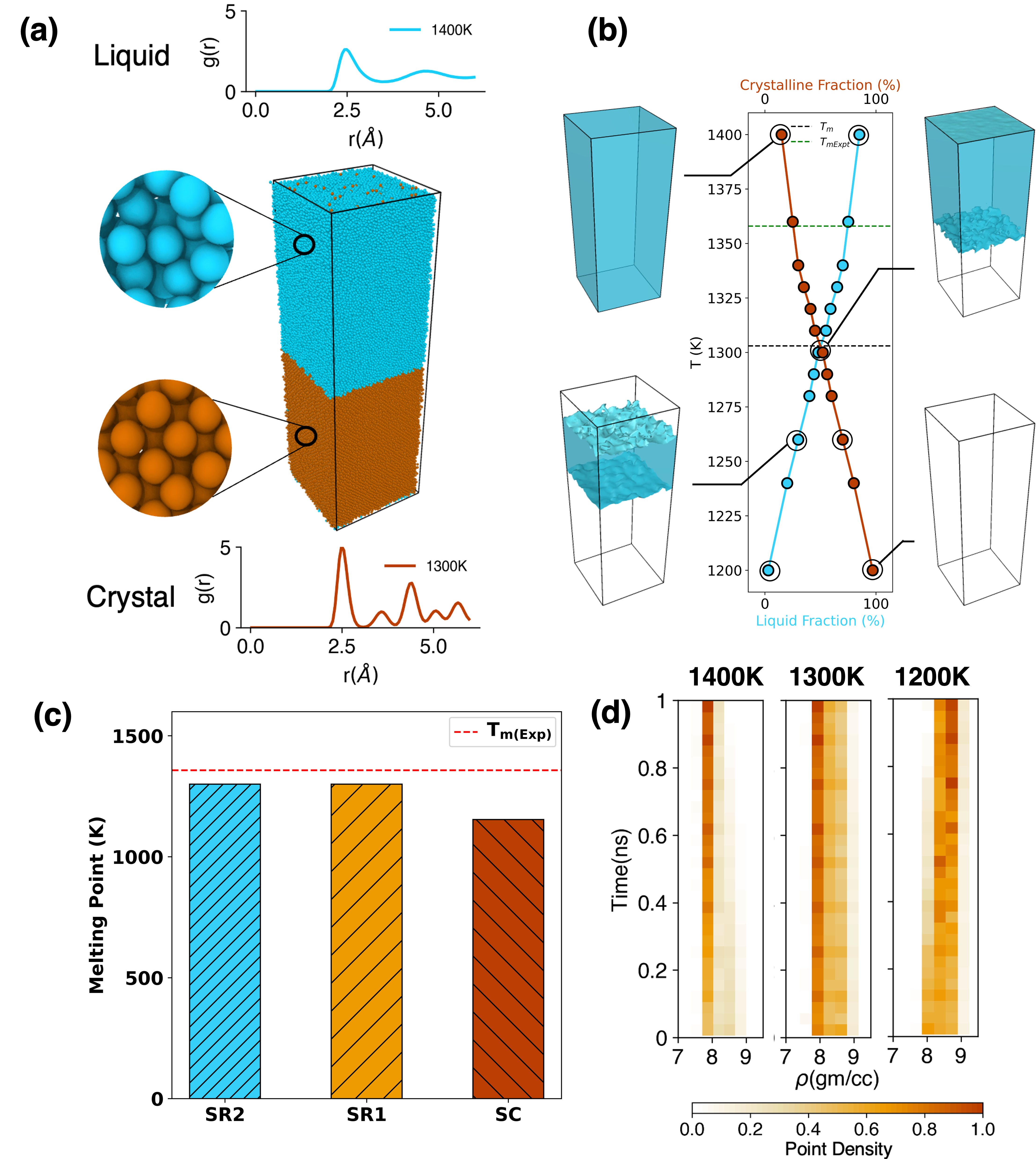}
\caption{Melting point calculations. (a) Initial two-phase simulation setup showing the solid-liquid interface and radial distribution functions (RDFs) for both phases. (b) Coexistence evolution curves for SR2 showing the relative amounts of solid and liquid after a \SI{1}{\nano\second} simulation at varying temperatures, with system snapshots at \SI{1200}{\kelvin}, \SI{1260}{\kelvin}, \SI{1300}{\kelvin}, and \SI{1400}{\kelvin}. The dashed black line marks the experimental melting point, and the dashed blue line indicates the SR2-predicted melting point. (c) Bar plot comparing melting point predictions from SR1, SR2, and Sutton-Chen EAM models. (d) Density evolution over time for simulations at \SI{1400}{\kelvin}, \SI{1300}{\kelvin}, and \SI{1200}{\kelvin}, showing phase transitions at each temperature.}
\label{fig:MeltingPointComparison}
\end{figure*}

The two-phase solid-amorphous interface simulation is a robust method for predicting the melting point in MD simulations by modeling the coexistence of crystalline and liquid phases at equilibrium. This approach avoids artifacts like superheating or undercooling and directly determines the melting temperature by adjusting conditions until no net phase change occurs. It provides a stringent test for validating interatomic potentials, ensuring they accurately capture the thermodynamic and dynamic properties of the solid-liquid transition. We therefore use a set of two-phase simulations (see Methods) to determine the melting points of copper for our SR1 and SR2 potential models and compare them to the melting point predictions of the SC-EAM model. We also compare them to the experimental melting point of Cu, which is \SI{1358}{\kelvin}. The melting dynamics and melting point analysis are illustrated in Fig.~\ref{fig:MeltingPointComparison}. Panel (a) shows the initial setup of the two-phase system, with a liquid and solid phase of copper in contact, accompanied by radial distribution functions (RDFs) of both phases to confirm their respective structures. Panel (b) presents the phase coexistence evolution curves for SR2, tracking the proportion of liquid and solid phases after a \SI{1}{\nano\second} simulation at various temperatures. This panel includes snapshots of the system at different temperatures (\SI{1200}{\kelvin}, \SI{1260}{\kelvin}, \SI{1300}{\kelvin}, \SI{1400}{\kelvin}), showcasing how the two-phase box evolves over time. The dashed black line indicates the experimental melting point, while the dashed blue line represents the melting point predicted by SR2. 

The SR1 and SR2 models yield an estimated melting temperature of approximately \SI{1300}{\kelvin}, which deviates by only 4.5 \% from the experimentally determined value of \SI{1358}{\kelvin}. The SC-EAM, however, yielded a significantly lower melting temperature of ~\SI{1154}{\kelvin}, indicating a limited ability to accurately model copper's melting behavior. Panel (c) displays a bar plot comparing melting point predictions made by SR1, SR2, and SC-EAM, visually emphasizing the accuracy of SR1 and SR2 in approximating the experimental value. Finally, panel (d) illustrates the density evolution over time for simulations at \SI{1400}{\kelvin}, \SI{1300}{\kelvin}, and \SI{1200}{\kelvin}. At \SI{1400}{\kelvin}, the system transitions from solid to liquid, indicating temperatures above the melting point; at \SI{1300}{\kelvin}, the system remains stable, suggesting near-equilibrium conditions; and at \SI{1200}{\kelvin}, the system transitions from liquid to solid, indicating temperatures below the melting point. Together, these results highlight the SR2 model’s high fidelity in predicting copper’s melting dynamics and phase transitions, making it a promising tool for simulating high-temperature applications and phase transition studies of copper.
\subsection*{Comparison of Models}

We have systematically developed and compared three distinct EAM-type potential models for copper:
\begin{enumerate}
    \item \textbf{SC EAM}: A Sutton--Chen--based embedded atom method that has historically offered computational simplicity and reasonable accuracy in bulk, near-equilibrium regimes.
    \item \textbf{SR1}: A symbolic regression--derived model in which the $\sqrt{\rho}$ embedding function is retained (mirroring the Sutton--Chen approach), but the pair and density functions are learned directly from DFT data.
    \item \textbf{SR2}: A more flexible symbolic regression model in which the pair, density, and embedding functions are allowed expanded functional forms, thereby capturing more complex many-body interactions.
\end{enumerate}

A fundamental benchmark of any interatomic potential is its \emph{equilibrium bond length}, \emph{binding energy}, and \emph{curvature} (the second derivative of the potential at the minimum). These reflect how atoms interact around their equilibrium separation and how stiff the potential well is near that point. Table~\ref{tab:EqProps} summarizes the key equilibrium properties for the three potentials:

\begin{table}[ht!]
\centering
\caption{Comparison of equilibrium properties (bond length, binding energy, and curvature) as predicted by SC, SR1, and SR2.}
\label{tab:EqProps}
\begin{tabular}{lccc}
\toprule
\textbf{Model} & \textbf{Bond Length} (\AA) & \textbf{Binding Energy} (eV) & \textbf{Curvature} (eV/\AA$^2$)\\
\midrule
SC EAM & 2.3501 & -1.1801 & 5.7638 \\
SR1    & 2.4096 & -1.0504 & 4.0343 \\
SR2    & 2.4138 & -1.0602 & 4.0663 \\
\bottomrule
\end{tabular}
\end{table}

\noindent
\textbf{SC EAM} (Sutton--Chen) predicts a notably shorter equilibrium distance, a deeper well, and higher stiffness than typical DFT or experimental references. This tends to overly constrain atomic displacements and can lead to significant errors for far-from-ground state configurations (e.g., defects, surfaces, or phases at high temperature).

Both \textbf{SR1} and \textbf{SR2} show equilibrium bond lengths and binding energies more aligned with DFT, reflecting improved transferability across a broader range of configurations. In particular:

\begin{itemize}
    \item \emph{SR1} retains the classical $\sqrt{\rho}$ embedding form but learns a more flexible pair and density term, thus outperforming SC EAM for various bulk, defect, and high-energy structures.
    \item \emph{SR2} extends beyond $\sqrt{\rho}$ by adding polynomial corrections (terms in $\rho$ and $\rho^2$). This augmented embedding term better captures many-body effects and non-equilibrium regimes, further improving surface and defect energetics, melting temperature predictions, and elastic constants.
\end{itemize}

Our results across bulk mechanical properties (elastic constants, phonon dispersion), melting behavior, and surface energies confirm these trends. While SC EAM matches some equilibrium properties reasonably, it shows significant deviations in, for instance, $C_{44}$, melting temperature, and high-index surface energies. In contrast, SR1 and SR2 accurately capture both low- and high-index surfaces, predict melting points within $\sim 4.5 \%$ of experiment, and yield more minor errors in forces and energies across diverse configurations. Notably, SR2 also yields improved $C_{44}$ and more reliable high-temperature phase predictions.

Overall, the more flexible \textbf{SR2} model provides superior accuracy and transferability for copper, without sacrificing the interpretability inherent in a symbolic regression framework. \textbf{SR1} remains a simpler alternative that already surpasses the classical SC EAM, but \textbf{SR2} stands out when modeling systems far from equilibrium. These findings underscore the power of symbolic regression in deriving next-generation interatomic potentials that are both physically interpretable and quantitatively robust for a wide range of thermodynamic and structural environments.

\begin{figure*}[htb]
\centering
  \includegraphics[width=1.0\textwidth]{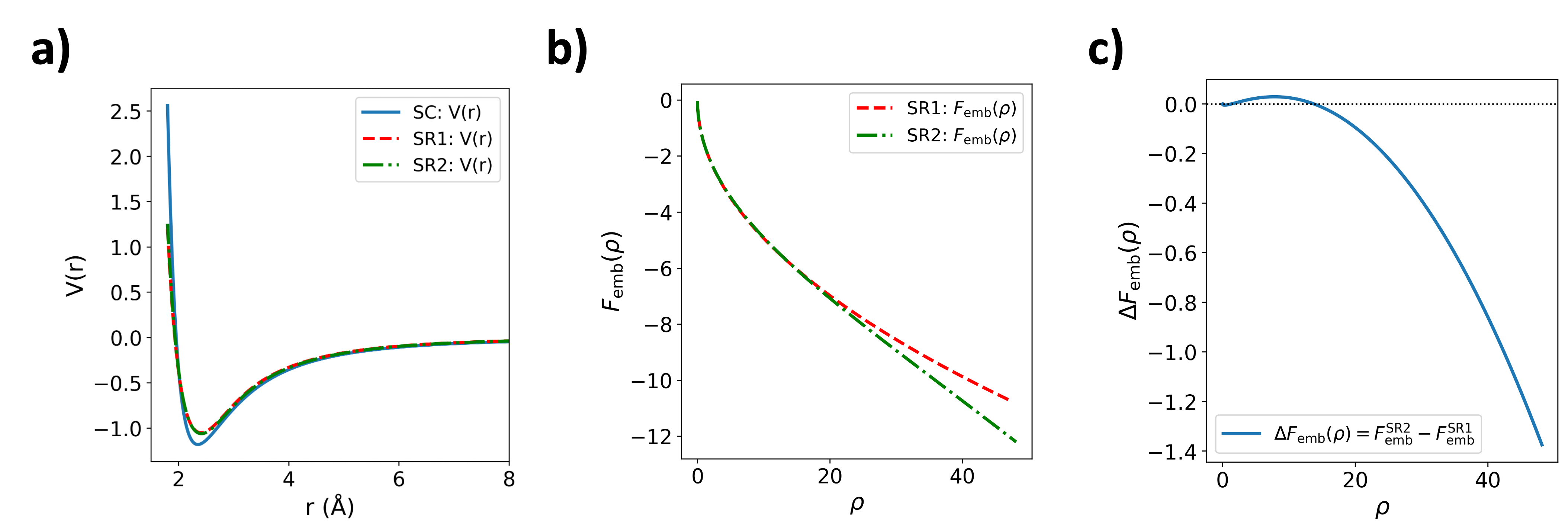} 
 
  \caption {Comparison of symbolic regression (SR)-derived potentials with the Sutton–Chen (SC) embedded-atom model (EAM). (a)Pair-wise interaction energies from SR2 (green), SR1 (red), and SC–EAM (blue) are plotted as a function of interatomic separation \(r\) (\si{\angstrom}). 
All three show the expected strong short-range repulsion and a single attractive well with \(V(r)\!\to\!0\) at large \(r\).
SR1 and SR2 nearly coincide across the bonding region, while SC–EAM is stiffer at short range and exhibits a slightly deeper well depth near the minimum. (b) Embedding energy \(F_{\mathrm{emb}}(\rho)\) as a function of electron density \(\rho\) for SR2 (green) and SR1 (red). 
The two models coincide at low densities and progressively diverge with increasing \(\rho\), with SR2 yielding a more attractive (more negative) embedding energy at high \(\rho\).
(c) Pointwise difference \(\Delta F_{\mathrm{emb}}(\rho)=F_{\mathrm{emb}}^{\mathrm{SR2}}(\rho)-F_{\mathrm{emb}}^{\mathrm{SR1}}(\rho)\), highlighting the density-dependent deviation (negative values indicate SR2 is deeper than SR1).
Together, the panels show that SR2 introduces a stronger embedding response at higher local densities while preserving agreement in the low-\(\rho\) regime.}
\end{figure*}

\subsection*{Influence of the Embedding Function on Many-Body Effects}
A key distinction between the SR1 and SR2 models lies in the form of their respective embedding functions, \(\,F_{\text{emb}}(\rho)\), which directly modulate the many-body interactions in EAM-type potentials. In \textbf{SR1}, the embedding function is restricted to a simple \(\sqrt{\rho}\) dependence, inherited from the Sutton--Chen formalism. While this captures a portion of the many-body physics, its functional flexibility is limited, particularly at higher electron densities. By contrast, \textbf{SR2} augments the \(\sqrt{\rho}\) term with \(\rho\) and \(\rho^2\) polynomial contributions, enabling more nuanced behavior over the entire density range. This added flexibility is especially relevant in regions of high coordination or near defect sites where the local electron density can deviate substantially from bulk equilibrium values. For instance, in under-coordinated surfaces or near vacancies, the local density profile may shift enough that the extra polynomial terms in \textbf{SR2} become crucial for accurately modeling the total many-body energy. At moderate to large densities (e.g., \(\rho \gtrsim 5\)), the additional terms in SR2 lead to increasingly negative embedding values, reflecting a stronger many-body stabilization effect. This shape difference manifests as improved performance in predicting surface energies and defect formation energies, where electron density variations are pronounced. Ultimately, the enhanced functional form in SR2 provides a superior match to DFT reference data across a broader range of local environments, underscoring the importance of embedding flexibility in capturing key many-body effects.

\subsection*{Model Interpretability}
While the symbolic expressions discovered by EqNN may become complex and less immediately intuitive than canonical EAM forms, they retain the fundamental advantage of explicit analytical transparency that distinguishes them from black-box neural network approaches. The EAM-like decomposition into pair interactions and many-body embedding terms ($E=\sum_{i<j}E_{\mathrm{pair}}(r_{ij})+\sum_i F_{\mathrm{emb}}(\rho_i)$) anchors interpretability in established physical concepts, allowing energies and forces to be traced to specific bonds and atomic environments. Unlike physics-informed neural networks or machine-learned interatomic potentials that embed their learned representations within thousands of interconnected weights and hidden layer activations, EqNN produces explicit closed-form analytic functions with a finite set of named, physically meaningful parameters that can be directly inspected, algebraically simplified, tabulated for rapid evaluation, and seamlessly deployed in existing atomistic simulation packages without any dependence on the original training architecture or specialized neural network libraries. Even when the resulting expressions involve higher-order corrections or nested exponentials that blur individual parameter meanings, the symbolic nature enables explicit identification of which aspects of the original EAM framework are insufficient and how they can be systematically improved. This approach successfully balances the accuracy demands of modern DFT-trained potentials with the computational efficiency and physical insight of traditional analytical models, offering a complementary pathway between the rigid constraints of classical EAM and the interpretability limitations of fully black-box approaches.
\subsection*{Model Performance - Computational Efficiency}
Table~4 benchmarks the computational efficiency of symbolic regression (SR) potentials in LAMMPS on a single CPU core (AMD EPYC 7763, 2.7~GHz), reporting wall-clock timing per atom per MD step. 
The SR potentials (SR1: 4.50~$\mu$s/atom/step, SR2: 4.41~$\mu$s/atom/step) are only marginally slower than the empirical Sutton--Chen EAM potential (3.60~$\mu$s/atom/step), while being two to three orders of magnitude faster than common machine-learning interatomic potentials (MLIPs). 
For example, GAP and MACE-Small require 1512 and 42~$\mu$s/atom/step, respectively, while SNAP and qSNAP both exceed 40~$\mu$s/atom/step, and MTP requires 20.7~$\mu$s/atom/step \cite{bernstein2024gap,jacobs2025practical}.
Thus, SR1 and SR2 achieve near-classical EAM performance while delivering accuracy comparable to DFT, providing a compelling balance of speed and fidelity. 
These results highlight that SR-based models offer a computationally tractable alternative to more expensive MLIPs, making them ideally suited for large-scale molecular dynamics simulations on leadership-class computing platforms. These results confirm that the SR models are only marginally slower than standard EAM, while maintaining their improved accuracy and interpretability.

\begin{table} [h]
\centering
\caption{Benchmarking computational performance of SR potentials in LAMMPS on a single core of an AMD EPYC 7763 (Milan) CPU at 2.7 GHz. Reported values are in $\mu$s/atom/step.}
\label{tab:performance}
\begin{tabular}{l c}
\hline
\textbf{Potential} & \textbf{Timing ($\mu$s/atom/step)} \\
\hline
SR1              & 4.50 \\
SR2              & 4.41 \\
Sutton--Chen EAM & 3.60 \\
MLIPs (GAP) & 1512 \cite{jacobs2025practical} \\
MLIPs (SNAP) & 40.9 \cite{jacobs2025practical} \\
MLIPs (qSNAP) & 41.3 \cite{jacobs2025practical} \\
MLIPs (MTP) & 20.7 \cite{jacobs2025practical} \\
MLIPs (MACE-Small) & 42 \cite{bernstein2024gap} \\
\hline
\end{tabular}
\end{table}

\section*{Discussions}

Physics-based models are traditionally employed in molecular simulations due to their foundation in established physical principles, offering robust and interpretable frameworks for understanding material behavior. The transition toward next-generation molecular simulation models hinges on the ability to overcome the constraints imposed by pre-defined functional forms, which often limit the accuracy and predictive power of these physics-based models. Integrating symbolic regression and reinforcement learning can overcome this limitation by improving accuracy and flexibility while retaining their interpretability, enabling the exploration of complex phenomena beyond traditional approaches. This work demonstrates the transformative potential of machine learning, specifically symbolic regression (SR), in directly learning the underlying multiscale physics of materials from data. By leveraging equation learner networks interfaced with reinforcement learning (RL) techniques such as continuous action Monte Carlo Tree Search and gradient descent, we develop a framework for deriving physically interpretable equations that accurately represent interatomic interactions.

A key feature of our training approach lies in the robust albeit relatively sparse dataset generated using the nested ensemble sampling method based on density functional theory (DFT). This dataset encompasses diverse configurations---ranging from crystalline to highly disordered and spanning ground state to far-from-ground state energy regimes---providing a comprehensive foundation for model training. Through rigorous benchmarking, our workflow successfully recovers known physical functional forms, such as those in the Sutton-Chen EAM formalism, while addressing its limitations. For representative transition metals such as Cu, we performed an unconstrained search of function space by incorporating mathematical operators, analytic functions, and material property variables to yield optimal models that minimize training error while enhancing interpretability.
A comprehensive performance assessment of the symbolic regression-derived models (SR1 and SR2) suggests that these models significantly outperform traditional EAM models in capturing key material properties and behavior, especially in regions far from equilibrium. These include equations of state, elastic constants, phonon dispersion, defect formation energies, surface/bulk energetics, as well as polymorph ordering and phase transformations. Notably, SR models also provide superior accuracy in representing melting dynamics—a stringent test of force field reliability. The two-phase solid-amorphous interface simulations reveal that SR models effectively capture the complex interplay of vibrational entropy, cohesive energy, and structural dynamics, achieving both quantitative and qualitative improvements over SC-EAM, when compared with experiments. Beyond accuracy, the interpretability of SR-derived equations is a substantial advantage, as it offers insights into the fundamental chemical trends and correlations inherent in the data. This level of understanding not only improves the reliability of the models but also opens avenues for guiding material design and discovery. The fast and accurate prediction capabilities of SR models, coupled with their physically meaningful representation of potential energy surfaces, mark a significant advancement in molecular simulations.

Overall, this work underscores the potential of symbolic regression and reinforcement learning to revolutionize the development of molecular simulation models. By providing fast, interpretable, and accurate representations of material behavior across spatiotemporal scales, the methods outlined in this work pave the way for model development that provides deeper insights into material dynamics and facilitates the design of advanced materials for a wide range of applications.

\section*{Methods}

The symbolic regression workflow utilizing equation learner networks\cite{martius2016extrapolation,sahoo2018learning,werner2021informed,chen2020learning}, as illustrated in Fig. 1, consists of several essential steps that together facilitate the discovery of precise and interpretable interatomic potentials. Firstly, the process begins with training and test data set generation using a nested sampling algorithm \cite{loeffler2020active,goldpaper,nielsen2013nested,varughese2024active} that iteratively explores the potential energy surface (PES). In the schematic shown in Fig. \ref{fig:Workflow}(a), the vertical axis represents the potential energy, while the horizontal axes denote the phase-space volume accessible to the system. The black dots scattered across the phase space symbolize the "live set" members, which are configurations currently under consideration. At each iteration, a dotted line highlights the energy contour, signifying the threshold below which configurations are accepted. This nested sampling method systematically explores different regions of the PES by progressively focusing on lower energy states, thereby efficiently mapping out the energy landscape and capturing the relevant configurational diversity needed for model training and testing. Next, Fig. \ref{fig:Workflow}(b) shows a reduced-dimensional representation of the diverse dataset that is constructed to capture key features relevant to energy landscape variations. High-dimensional datasets from ab initio calculations contain vast amounts of information, much of which may be redundant or irrelevant for the potential model.

By employing dimensionality reduction techniques such as principal component analysis (PCA)\cite{abdi2010principal}, we map the essential configurational features influencing the energy variations. This simplified representation retains the critical physical insights and demonstrates the diversity of the configurations in our sampled training and test data set. A key step in our workflow is shown in Fig. \ref{fig:Workflow}(c), which depicts the Equation Neural Network (EqNN) architecture, where each node corresponds to a distinct basis function contributing to the overall model that needs to be learnt. The EqNN combines neural network capabilities with symbolic regression by representing the potential energy as a composition of mathematical expressions (basis functions). Each node in the network symbolizes a specific mathematical operation or function, such as addition, multiplication, or trigonometric functions. The network architecture allows for flexible combinations of these basis functions, enabling the construction of complex equations that accurately describe the interatomic potentials. This symbolic representation ensures that the resulting model is interpretable, as it can be expressed in terms of fundamental mathematical relationships. Finally, Fig. \ref{fig:Workflow}(d) depicts the overall training process, which involves iterative adjustments to optimize the network and improve model accuracy. During training, the EqNN adjusts the parameters and structure of the network to minimize the difference between the predicted and actual energies and forces from the training data.

We utilize a two-prong optimization procedure to derive the model. First, we use a derivative-free optimizer such as Monte Carlo Tree Search (MCTS)\cite{patra2020accelerating,manna2022learning} to perform a global search, followed by a gradient descent\cite{gilbert1992global} for more local search. MCTS is a heuristic global search algorithm that combines random sampling and strategic exploration to make decisions in large or complex search spaces. It builds a search tree incrementally through four steps: selection of promising nodes, expansion of the tree, simulation of outcomes via random rollouts, and backpropagation of results to update node values. As a reinforcement learning approach, MCTS learns optimal decisions by balancing exploration of new actions and exploitation of known good actions, using methods like the Upper Confidence Bound for Trees (UCT). The local search involves backpropagation and gradient descent techniques to fine-tune the weights and biases associated with each node. The iterative optimization continues until the model achieves a satisfactory level of accuracy, balancing complexity with flexibility. The result is a robust interatomic potential model that not only fits the training data well but also maintains physical interpretability through its symbolic mathematical form. This comprehensive workflow leverages advanced reinforcement learning algorithms such as MCTS and neural network architectures to perform symbolic regression, ultimately discovering interatomic potentials that are both accurate and physically meaningful. By integrating nested sampling, dimensionality reduction, and equation-learning neural networks, our approach addresses the challenges of discovering functional forms capable of modeling complex atomic interactions in materials. Each of the major steps in our workflow is illustrated in more detail below:

\begin{figure*}[ht]
\centering
\includegraphics[width =1.0\textwidth]{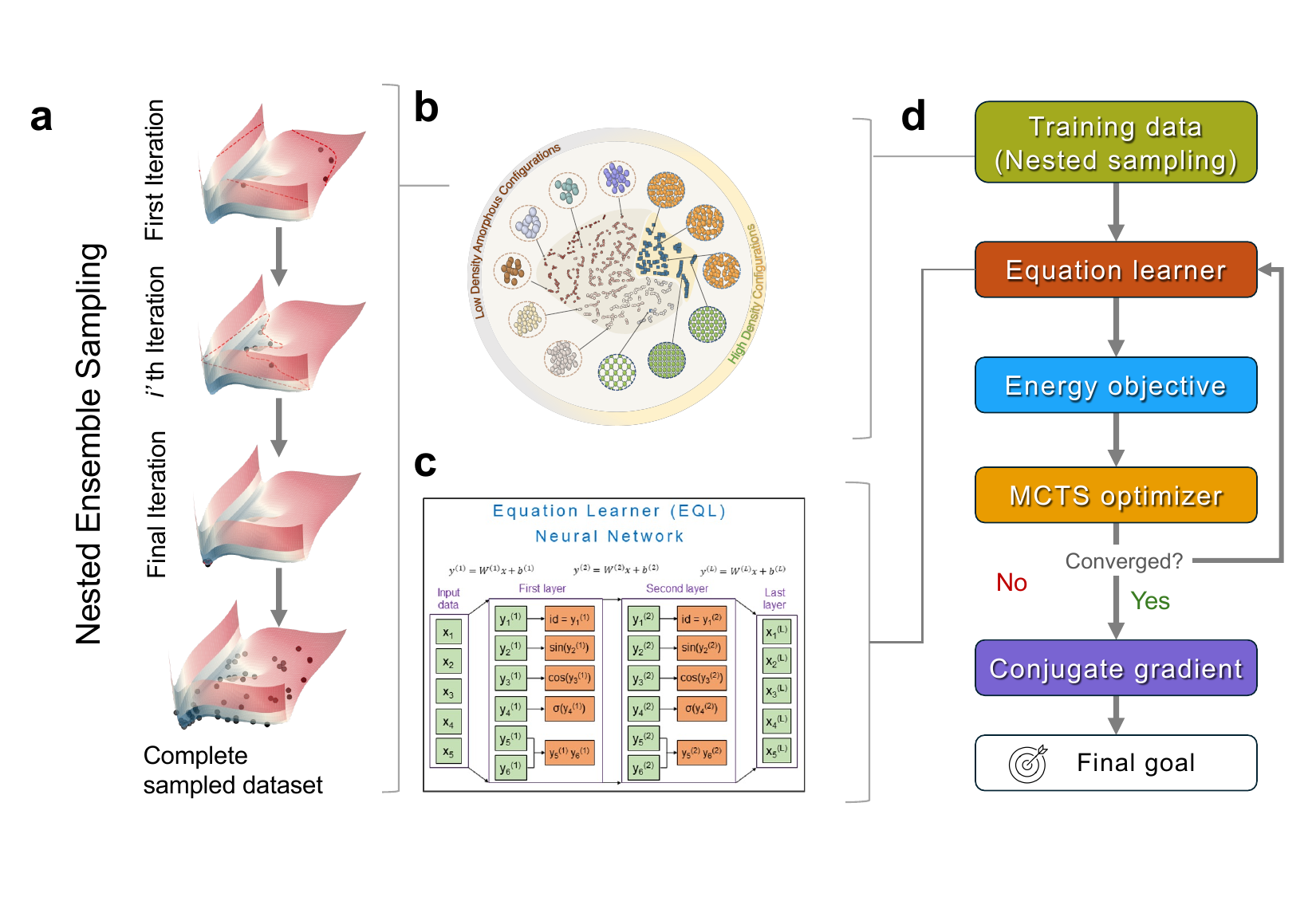}
\caption{The Workflow for Symbolic regression. (a) Schematic of the nested sampling algorithm illustrating iterative exploration across the potential energy surface (PES). The vertical axis indicates potential energy, while the horizontal axis represents phase-space volume. Black dots mark the live set members distributed across phase space, with a dotted line highlighting the energy contour at each step. (b) Reduced-dimensional representation of the training-set latent space (SOAP descriptors), highlighting structural families and energy-landscape variations (c) Diagram of the EqNN neural network architecture, where each node corresponds to a distinct basis function that contributes to the overall model. (d) Visualization of the training process, showing iterative adjustments to optimize the network and improve model accuracy.}
\label{fig:Workflow}
\end{figure*}

\subsection*{Equation Learner Neural Networks}

Equation learner networks are advanced frameworks that combine the principles of neural networks and symbolic regression to derive mathematical equations directly from data. These networks aim to uncover interpretable relationships by representing equations as compositions of mathematical operations and basis functions. Unlike traditional machine learning models that function as black boxes, equation learner networks prioritize interpretability by constructing equations in symbolic form. Among the several equation generation options \cite{kenoufi2015symbolic,hernandez2019fast,hernandez2023generalizability,wang2024exploring,schmidt2009distilling,udrescu2020ai}, equation learner networks have been explored as a potential way to determine equations using more traditional neural network training approaches.  

In the networks employed in this work, the normal activation functions found in common neural networks are replaced with functional activation representing various mathematical operations.  Instead of sigmoids, rectified linear units, or other similar activation functions, elementary operations such as reciprocal functions, squares, etc.---called basis functions as they make up the various elementary functional forms---are used. Hence, the resulting model, when trained, can be collapsed into a single mathematical equation by the end. The upside of these approaches is that one may back-propagate the coefficients of the matrix to determine the weights, and as such, remain compatible with various gradient-based approaches. However, a major downside is that, unlike traditional neural networks, where the functions are explicitly chosen for their stability, the functions in these symbolic networks have more problems associated with their use. 

The architecture of equation learner networks includes nodes representing distinct mathematical functions, such as addition, multiplication, or more complex operators like trigonometric or exponential functions. These nodes are interconnected to form a flexible structure capable of representing a wide range of equations. The network starts with an initial set of randomly generated equations and iteratively refines them through optimization. This optimization minimizes the error between the predictions of the derived equation and the actual values in the training dataset. By allowing both the structure and parameters of the equations to evolve, equation learner networks can adapt to the specific complexities of the data they analyze.

A key feature of equation learner networks is their ability to perform an unconstrained search over the function space. Instead of relying on predefined functional forms, the networks explore potential solutions by combining mathematical operators and constants dynamically. This flexibility enables them to discover novel equations that capture complex relationships in our training data. The iterative training process, described in more detail in the next sections, involves fine-tuning the equations to accurately reproduce target properties, such as energies, forces, or stresses, while maintaining simplicity and interpretability. This balance between accuracy and comprehensibility is critical for applications requiring both predictive power and physical insight.

By leveraging these networks, we identify physics-based fast, accurate, and transferable potential models directly from ab initio data. The resulting equations not only enable efficient simulations over large time and length scales but also provide physical insight into the governing interactions. This combination of interpretability, adaptability, and accuracy makes equation learner networks a powerful tool for advancing our understanding of complex systems and bridging the gap between empirical and machine learning-based approaches.

\subsection*{Nature of the Datasets - Training, Validation and Test Data Generation}

We employed two sets of data. (i) We generated training and test data using an active learning approach to sample configurations for benchmarking the EqNN by recovering the known functional forms (SC-EAM), and (ii) the energies of the sampled configurations in (i) were recalibrated with DFT calculations and used to generate new SR models that work well across a broad range of energies, from near to far-from--ground state.

\paragraph{Benchmarking data set for recovering SC-EAM model}

An active learning approach was employed to sample the training dataset. A key objective is to minimize the amount of training data required, particularly if the reference model is based on computationally expensive first-principles calculations. To achieve this, we deploy an active learning (AL) framework that trains the EqNN model dynamically, using a minimal amount of training data. For benchmarking, the data are generated using the SC-EAM model. The AL workflow begins with a sparse initial dataset (approximately 21 data points) and is updated iteratively through a Nested Ensemble Monte Carlo approach. This method identifies regions of the energy landscape where the model performs poorly, selectively queries these regions, and updates the training pool to enhance the network's accuracy and performance. Initially, a small set of training structures, representing configurations near ground state FCC and their corresponding energies were computed. A preliminary potential model was trained to capture the energies of this initial structure set with reasonable accuracy. Subsequently, this potential model facilitated the generation of new structures via nested ensemble sampling. The predicted energies from the current model state were then compared with those derived from the reference SC-EAM. For each sampled structure in the test set, if the symbolic neural network and the EAM prediction do not agree within a given tolerance (which was 110 percent of the training error for the current state of symbolic NN for this work), the structure is then added to the training pool. And the symbolic NN model is trained once again with the updated training dataset.  This iterative process of model retraining and structure generation continued until a model closely approximating, or equivalent to, the original Sutton-Chen equation was achieved. At the end of this active learning cycle, a total of 341 structures were generated and found to be sufficient to recover a functional form close to that of the original SC-EAM.

\paragraph{Training data set for new symbolic regression (SR) models}

We utilize the final training set obtained from stage one (where benchmarking of the EqNN is performed by recovering a known SC-EAM model) and recalculate their energy values using Density Functional Theory (DFT)\cite{hohenberg1964inhomogeneous} using the Vienna Ab initio Simulation Package (VASP)\cite{kresse1996efficiency}. This refined dataset serves as the foundation for retraining our EqNN to derive new SR models, with a particular emphasis on minimizing the mean absolute error in energy prediction for both the training and test sets, rather than deriving specific equations. As discussed later, our training approach incorporates a weighted mean absolute error loss function, assigning greater weight to structures near equilibrium—those representing the equations of state compared with those farther from equilibrium. This strategy enhances the accuracy of predictions around the most critical regions.

\paragraph{Nested ensemble for configuration sampling}

In our active learning workflow, the sampling of the configurations is performed using the nested ensemble sampling method. This approach efficiently samples structural configurations across the different energy ranges, providing a robust dataset for model training. We begin by generating a large number of initial configurations of the system, ensuring that these configurations uniformly cover the entire phase space. In a system consisting of N particles, random positions were generated for each particle within the defined boundaries. During the initial sampling procedure, all generated configurations were accepted regardless of their energies. The potential energy E for each configuration was computed using the relevant potential energy function, i.e., the current state of the symbolic NN being trained. The energies of all accepted configurations were recorded, and the median energy E$_1$ was calculated from these values. This median energy served as the threshold for the next iteration. The configuration and corresponding energy information were stored in the structure set, ensuring a comprehensive record of the initial sampling phase.

For the next iteration, the median energy from the first iteration, E$_1$, was used as the energy threshold. New configurations were generated by perturbing the accepted configurations from the previous iteration, and only those with energies lower than E$_1$ were accepted. The energies of these configurations were recorded, and a new median energy E$_2$  was computed. This iterative process continued, with each iteration using the median energy from the previous iteration as the new energy threshold. For each iteration n, the median energy E$_{n-1}$ was recorded along with the configuration and energy values. The final structure set was compiled by combining the configuration and energy data from all iterations.  This compilation ensured that the structure set included a diverse range of configurations with corresponding energy values covering the relevant configuration space comprehensively.

\subsection*{EqNN Architecture}

A typical requirement for symbolic neural networks is that the model architecture must often contain a sizable number of terms in its set of basis functions to be flexible enough to be able to identify a usable equation.  As such, many of these functions must be fixed prior to running. Because of this, there needs to be a consistent way to eliminate functions in the network that are not useful for fitting an equation to a curve.  This is less straightforward in a symbolic network containing mathematical functions that are significantly different from each other.   Traditionally, in neural networks, setting the weights associated with a given node to a small value would turn off that node, such that it no longer contributes to the network.   However,  many functions, even with small coefficients, can still have very large contributions.  For example, if one has a reciprocal function, then the gradient as the weight approaches zero will be infinite for any weight besides zero.  This can result in situations where a small value for a given weight will have a very large contribution to the final output of the network as the inputs approach 0. Traditional approaches, such as simply removing functions whose weights are very small, can potentially be wrong because their contribution might still be required. As such, it is more desirable to switch weights on and off during training to let the optimizer decide which weights to use. To do this, a pseudo-weight was defined.  For a given weight in the symbolic network, denoted $w_{i}$, a corresponding pseudo-weight, $v_{i}$, is created. These pseudo-weights are what the optimizer will directly perform the optimization on.  

\[
    w_{i}= 
\begin{cases}
    v_{i} + L_{zone} &  v_{i} \leq  -L_{zone} \\
    0,           & -L_{zone} \leq v_{i} \leq L_{zone}\\
    v_{i} - L_{zone},&   L_{zone} \leq v_{i} 
\end{cases}
\]

Where $L_{zone}$ corresponds to the length of the null zone.  This is a user-specified parameter.  The larger the width of $L_{zone}$, the more aggressively the optimizer will turn off weights in the network.  If $L_{zone}=0$, then the pseudo-weights will be exactly equal to the real weights.  The purpose of this is to increase the chance that the optimizer will set a weight exactly to 0.   Under normal circumstances, the optimizer can often get very close to 0, but it is nearly impossible to reach exactly 0.  By using these pseudo-weights, the optimizer has a wide zone that it can use to turn off weights during the optimization. It is worth noting that if $L_{\text{zone}}$ is chosen too large, basis functions with significant contributions may be prematurely eliminated; in such cases, the prediction error can increase sharply, and re-optimization of the remaining parameters may not stabilize the model. For this reason, we fixed $L_{\text{zone}} = \pm 0.125$, which we found to balance pruning efficiency with training stability.

Another problem one encounters is that, unlike normal activation functions that have been chosen for their stability in both their values and gradients, the various basis functions can have very different mathematical properties, which can yield unruly results.  It is possible for some of the basis functions of interest to have domain-related errors that can result in NaN or infinity occurring in the network.   While one can adjust the starting condition of the network to prevent the gradient from running into such a situation, it is difficult to iterate over the entire search space with gradient-based approaches alone while preventing this problem.  However, for interatomic potentials, these function types are important due to the natural requirement that the interaction energy between a group of atoms drops off as the distance grows larger.  It is possible to assign a large value as a penalty in the loss function during the optimization to prevent overflow and other similar errors.  This also prevents the optimizer from crashing during a run.

\subsection*{The EqNN-based Networks}
The EqNN architecture was designed to mimic the embedded atom method (EAM) and represent the energetic interactions of the various sampled atomic configurations. In the EAM model, the total energy is represented by the following equation:

\begin{equation}
E_{\text{total}} = \sum_i \left[ \frac{1}{2} \sum_{j \neq i} \phi(r_{ij}) + F\left(\sum_{j \neq i} \rho(r_{ij})\right) \right]
\end{equation}\cite{sutton1990long}

In the above equation,
\begin{itemize}
    \item \( E_{\text{total}} \) is the total energy of the system.
    \item \( \phi(r_{ij}) \) is the pair potential between atoms \( i \) and \( j \).
    \item \( F(\rho_i) \) is the embedding function which depends on the electron density \( \rho_i \) at atom \( i \).
    \item \( \rho(r_{ij}) \) is the contribution to the electron density at atom \( i \) due to atom \( j \).
    \item \( r_{ij} \) is the distance between atoms \( i \) and \( j \).
\end{itemize}

We retain the core physics represented by the EAM-type models and hence, our EqNN architecture consists of three neural networks trained simultaneously, each responsible for different components of the EAM, ie, pair, density, and embedding functions.
The EqNN model is initialized with a set of activation functions, a cutoff radius, a maximum number of atoms, and an offset parameter.

The three neural networks in the EqNN are:
(i) \textbf{Pair Potential Network} which computes the energetic interactions between pairs of atoms.
(ii) \textbf{Density Network} which computes the local electron density contributed by neighboring atoms.
(iii) \textbf{Embedding Energy Network} which computes the many-bodied energy contribution as a function of the local electron density. Each neural network has a specified number of layers and activation functions designed to capture the complex interactions characteristic of the EAM potential. The activation functions used in the model are critical for accurately representing the EAM potential as they form the basis functions for the search space, and the final equation is represented as a combination of these functions.

The chosen functions for each component are as follows: (i)\textbf{Pair Potential}: Includes inverse powers and negative exponentials to capture the distance dependence of the pair interactions. (ii) \textbf{Electron Density}: Includes inverse powers and negative exponentials to model the contribution of neighboring atoms to the local electron density. (iii) \textbf{Embedding Energy}: Includes square roots, squares, and higher-order polynomials to capture the nonlinear dependence of the embedding energy on the local electron density.

\begin{table}[h!]
\caption{Basis Functions for Pair, Density, and Embedding used in our EqNN}
\centering
\begin{tabular}{|c|c|}
\hline
\textbf{Function Type} & \textbf{Basis Functions} \\ \hline
\multirow{9}{*}{\textbf{Pair}} 
  & $\exp(-r)$ \\ \cline{2-2} 
  & $r^{-1.0}$ \\ \cline{2-2} 
  & $r^{-6.0}$ \\ \cline{2-2} 
  & $r^{-7.0}$ \\ \cline{2-2} 
  & $r^{-8.0}$ \\ \cline{2-2} 
  & $r^{-9.0}$ \\ \cline{2-2} 
  & $r^{-10.0}$ \\ \cline{2-2} 
  & $r^{-12.0}$ \\ \cline{2-2} 
  & $r^{-14.0}$ \\ \hline
\multirow{6}{*}{\textbf{Density}} 
  & $\exp(-r)$ \\ \cline{2-2} 
  & $r^{-1.0}$ \\ \cline{2-2} 
  & $r^{-5.0}$ \\ \cline{2-2} 
  & $r^{-6.0}$ \\ \cline{2-2} 
  & $r^{-7.0}$ \\ \cline{2-2} 
  & $r^{-8.0}$ \\ \hline
  \multirow{1}{*}{\textbf{Embedding - SR1}} 
  & $\rho^{0.5}$ \\ \hline
\multirow{5}{*}{\textbf{Embedding - SR2}} 
  & $\rho^{0.5}$ \\ \cline{2-2} 
  & $\rho^{1}$ \\ \cline{2-2} 
  & $\rho^{2}$ \\ \cline{2-2} 
  & $\rho^{3.0}$ \\ \cline{2-2} 
  & $\rho^{4.0}$ \\ \hline
\end{tabular}

\label{tab:basis_functions}
\end{table}
We train two types of EqNN (SR1 $\&$ SR2) that generate symbolic models with varying degrees of flexibility. In the first (SR1), we constrain the basis functions for the embedding term solely to the square root function, mirroring the embedding term in a typical Sutton-Chen EAM formalism. Note that the pair-wise and the density networks are allowed to explore an expanded set of basis functions. 
The EqNN model was trained using Monte Carlo Tree Search (MCTS), followed by gradient descent until the loss function achieved the desired threshold of less than \SI{20}{\milli\electronvolt\per\atom}.

In the second (SR2), we allow an expanded basis set even for the embedding term to include a variety of functions, leveraging existing knowledge from the Sutton-Chen EAM formalism while aiming to further enhance its predictive capabilities in far-from-ground state regions. Specifically, we broadened the basis set for the embedding term to encompass various polynomial functions. To capitalize on the progress made in the first approach, we used the best-performing network from the initial method as the starting point for training in this expanded basis set approach.

The input data to the EqNN is a tensor representing the distance matrices of atomic configurations, and an array specifies the number of atoms in each structure. The distance matrix tensor has elements r$_{ij}$ for distances between atoms i and j if the distance falls within the cutoff for the potential model and zero otherwise. It is padded with zeros for pairs of atoms that do not exist for structures with atoms less than N$_{max}$ to ensure consistent dimensions across different structures. The distance matrix tensor has dimensions (M * N$_{max}$), where M is the total number of atoms across all structures, and N$_{max}$ is the number of atoms in the largest structure in the dataset. 

The EqNN workflow initiates with the computation of the density function for each atomic pair via the density network, ensuring values beyond the cutoff distance are zeroed by subtracting the density function at the maximum cutoff distance. Subsequently, the embedding energy calculation involves summing the density contributions for each atom, followed by inputting the resultant sum into the embedding energy network. This step generates the embedding energy, which is then reshaped to align with the input data structure. The subsequent pair potential calculation uses the pair potential network to compute the pair potential for each atomic pair, ensuring contributions beyond the cutoff are zeroed by subtracting the pair potential at the maximum cutoff distance. The energy summation step then aggregates the pair potential and embedding energy contributions for each structure, normalizing the total energy by the number of atoms in each structure. Consequently, the model outputs the total energy for each atomic configuration, normalized by the number of atoms, thus providing a comprehensive energy profile for each structure.

Both trained models were rigorously validated against energy and force predictions, as well as various structural properties, including the equation of states (EOS), phonon dispersion, elastic constants, vacancy formation energies, lattice constants, and cohesive energies. The results from these models were compared to corresponding Density Functional Theory (DFT) predictions, demonstrating a high degree of agreement between the machine learning models and the DFT calculations.

These findings underscore the effectiveness of EqNN in enhancing the predictive accuracy of our potential models, while also showcasing the potential of machine learning techniques to refine and expand upon traditional EAM formalisms. The comparative analysis between the models and DFT predictions validates the robustness and reliability of our methods in capturing essential structural properties, thereby advancing the practical application of machine learning in computational materials science.

\subsection*{Loss Function}
Development of an accurate symbolic model requires a comprehensive loss function that integrates constraints imposed by basis set functions and physical principles. Traditional loss functions often fail to account for domain-specific errors and physical constraints, leading to suboptimal model performance and potential numerical instabilities. Our modified loss function explicitly incorporates constraints arising from the basis set functions. These constraints are crucial for maintaining the physical realism of the model. In scenarios where the basis functions produce domain-related errors resulting in NaN or infinite values, the loss function assigns a significantly large penalty. This approach prevents the optimizer from crashing and ensures the stability of the training process. The EAM model necessitates that the pair and density functions yield positive outputs at all distances within the cutoff range. To enforce this, the loss function imposes a large penalty whenever any network outputs a negative value within the specified cutoff distance. This constraint is vital for maintaining the physical integrity of the model, as negative values could lead to non-physical predictions. Empirical observations and theoretical insights indicate that structures near their equilibrium configurations are more prevalent within the temperature ranges of interest. These structures also play a critical role in determining properties such as elastic constants. Therefore, our loss function assigns higher weights to these near-ground state structures. This weighting scheme enhances the model's predictive efficiency, particularly for properties that are sensitive to structural variations. To quantify the accuracy of energy predictions, we use the mean absolute error (MAE) as our primary metric. Our comprehensive loss function is formulated as follows:

\begin{equation}
\text{Loss Function} = w \cdot \text{MAE}_{\text{NE}} + \text{MAE}_{\text{RE}} + \text{Penalty}_{\text{Pair}} + \text{Penalty}_{\text{Density}} + \text{Penalty}_{\text{Domain}}
\end{equation}
\[
    Penalty_{Pair}= 
\begin{cases}
    1e18 &  Pair(r) < 0 \\
    0,           &  Pair(r) > 0 
\end{cases}
\]
\[
    Penalty_{Density}= 
\begin{cases}
    1e18 &  Density(r) < 0 \\
    0,           &  Density(r) > 0 
\end{cases}
\]
\[
    Penalty_{Domain}= 
\begin{cases}
    1e18 &  NaN/Inf \ outputs \\
    0,           &  otherwise
\end{cases}
\]
Where:
\begin{itemize}
    \item $w$ is the weighting factor for near-ground state structures,
    \item $\text{MAE}_{\text{NE}}$ is the mean absolute error for near-ground state structures,
    \item $\text{MAE}_{\text{RE}}$ is the mean absolute error for the rest of the structures,
    \item $\text{Penalty}_{\text{Pair}}$ is the penalty for the pair function producing negative values within the cutoff,
    \item $\text{Penalty}_{\text{Density}}$ is the penalty for the density function producing negative values within the cutoff,
    \item $\text{Penalty}_{\text{Domain}}$ is the penalty for domain errors in the basis set functions.
\end{itemize}

\subsection*{EqNN Training - Global and Local Search}

\paragraph{Global optimization}

First, we optimize the weights of the EqNN using a global search algorithm. Monte Carlo Tree Search (MCTS) is a powerful global optimization algorithm that integrates probabilistic and heuristic methods, originally developed for discrete decision-making problems such as board games like Go and chess. The algorithm explores potential decision paths using a tree structure, alternating between exploration (investigating unvisited branches) and exploitation (focusing on branches with the highest rewards). The classical MCTS process involves four steps: (1) Selection, where nodes are ranked using selection rules like the Upper Confidence Bound (UCB); (2) Expansion, which generates child nodes by performing valid moves; (3) Simulation, or "playout," where random moves are performed until an outcome is achieved; and (4) Backpropagation, where the reward is propagated back to parent nodes. This process allows MCTS to balance computational efficiency and decision accuracy, making it effective for problems with large but discrete action spaces.
In continuous action spaces, traditional MCTS faces challenges due to the infinite number of potential child nodes, requiring adaptations to make it viable. Recent advancements, such as the development of a continuous-action MCTS (c-MCTS) \cite{manna2022learning}, redefine nodes to represent regions in a continuous search space, typically modeled as hyperspheres around parameter sets. The decision tree in c-MCTS refines the search space by combining random sampling (playouts) within these regions with an adaptive trade-off mechanism between exploration and exploitation. This ensures the algorithm efficiently identifies optimal solutions while avoiding local minima. Furthermore, c-MCTS’s ability to quickly grow new branches when trapped in metastable states makes it a robust tool for complex optimization tasks.
To address challenges in continuous action space problems, we have made modifications to c-MCTS by introducing three key innovations: (1) a uniqueness function to avoid degeneracy by ensuring distinct branches do not converge to the same solution; (2) correlating tree depth with the search space, progressively narrowing the search radius of child nodes relative to parent nodes for a more structured exploration; and (3) adaptive sampling of playouts, where random simulations are biased toward regions closer to the parent node to improve sampling efficiency in high-dimensional spaces. These advancements enable c-MCTS to effectively handle continuous, high-dimensional optimization problems, making it applicable to areas like physical simulations and material discovery.

In the symbolic regression workflow, the selection of hyperparameters plays a pivotal role, particularly the definition of the search space and the implementation of depth scaling. The continuous-action Monte Carlo Tree Search (c-MCTS) algorithm is employed to navigate the hyperspace of weights and biases for the Equation Neural Network (EqNN), which subsequently determines the coefficients in the final analytical expressions. A carefully calibrated search space is essential: it must be sufficiently expansive to capture relevant equations yet constrained enough to maintain computational efficiency. For the case of copper, we employed a search space bounded by upper and lower limits of ±5.0 for all parameters. These bounds were chosen to balance exploration and computational tractability, but their adaptation is advisable when applying the workflow to other systems. The depth scaling mechanism incorporates an adaptive geometric scaling rate of 0.5, with a depth limit set at 25 and a maximum search radius of 5.0. This configuration enables the algorithm to initiate with broader explorations of the parameter hyperspace, identifying promising regions before progressively refining its focus. As the search proceeds, the scaling of the search window ensures incremental optimization of the parameter sets, allowing the algorithm to systematically converge on high-quality solutions. For this symbolic regression workflow, we use a playout value of 10 that works well. 
We observe that the global search phase performs exceptionally well during the initial stages of the workflow, yielding high-quality equations and optimizing the loss function to within a few hundred meV/atom. However, as the search progresses, convergence becomes increasingly challenging. This deceleration arises because the algorithm continues to sample points randomly, even within the increasingly refined search space. While this random exploration supports broader search capabilities, it becomes less effective for fine-tuning solutions as the optimization landscape narrows.

To address this, we employ a transition from the global search algorithm to a local optimization approach, such as the Adam optimizer, at this critical juncture. This shift facilitates rapid convergence to higher-quality solutions, effectively reducing the loss function to within a few tens of meV/atom. By leveraging the strengths of global exploration in the early stages and localized refinement in the later stages, this hybrid optimization strategy ensures both the breadth and precision required for generating robust symbolic regression models. For this workflow, we use $\sim$ \SI{150}{\milli\electronvolt\per\atom} as a cutoff where we switch from global to a local optimizer.

\paragraph{Local optimization}
Gradient descent is next used in Equation Neural Networks (EqNN) to optimize the network parameters by minimizing the error between predicted and target values. Once the MCTS has gotten close enough to a reasonable set of weights or parameters, we perform a gradient descent to further relax the weights. We consider $~$\SI{100}{\milli\electronvolt\per\atom} training error with MCTS to be a good spot to start the gradient descent when we do not see an improvement in the training error for five thousand MCTS steps. This is done directly on the real weights instead of the pseudo-weights, and any weights turned off are excluded from the gradient.  During training, the network calculates gradients of the loss function with respect to its parameters, which indicate the direction and magnitude of changes needed to reduce the error. The gradients derived from the loss function are subsequently employed to iteratively update the parameters, leveraging adaptive optimization methods such as the Adam optimizer. In this study, we utilize a learning rate of $10^{-4}$, which strikes a balance between rapid convergence and stability, effectively minimizing oscillations in the loss function. The Adam optimizer proves particularly effective when initialized from a high-quality starting point near the optimal solution, as the gradients in such regions tend to be smooth and exhibit low noise. However, in scenarios where optimization begins from a random initialization within a high-dimensional search space characterized by a complex and rugged loss landscape, the Adam optimizer is prone to becoming trapped in local minima. To address this limitation, our workflow employs a hybrid strategy that synergistically combines the strengths of Monte Carlo Tree Search (MCTS) and the Adam optimizer. The global exploratory capabilities of MCTS are leveraged to navigate the hyperspace and identify promising regions, while the Adam optimizer refines these solutions by efficiently converging to high-quality minima. This hybrid approach ensures robust and efficient identification of optimal equations for modeling the material's potential energy landscape.
This process enables EqNN to fine-tune both the structure and coefficients of its symbolic representations, ensuring accurate and interpretable equations for modeling complex relationships such as interatomic potentials.

\subsection*{Melting point from two--phase coexistence}
We determined the equilibrium melting temperature $T_\mathrm{m}$ using a standard two-phase (solid-liquid) coexistence protocol in LAMMPS. The workflow was split into two stages: 1) construction of the solid--liquid system and 2) coexistence runs and detection of interface drift.

\paragraph{Initial two-phase configuration }
Starting from the relaxed crystalline phase, a simulation cell was replicated to obtain a slab that is sufficiently wide in the directions parallel to the interface and elongated along the interface normal (here, the $z$ axis) to host two independent solid--liquid interfaces under periodic boundary conditions. The cell was then partitioned into two half-spaces: a \emph{solid} region and a \emph{melt} region. The melt region was driven to the liquid by heating it well above an initial estimate of $T_\mathrm{m}$ (around \SI{2500}{\kelvin})under a thermostat, while the solid region was restrained to maintain crystallinity (We set the forces for all atoms in the crystalline region to be zero for this step). After complete melting, the liquid half was brought back (quenched) to a trial temperature $T_{\mathrm{trial}}$ close to the expected $T_\mathrm{m}$ to minimize density mismatch at the interface. The resulting solid-liquid bi-phase system (two planar interfaces normal to $z$) was briefly relaxed to remove residual stresses.

\paragraph{Coexistence simulations}The simulation box consisted of crystalline and amorphous slabs stacked along the z-direction, with periodic boundary conditions applied in all three spatial dimensions. Initial velocities were assigned from a Maxwell–Boltzmann distribution corresponding to the target temperature, and the total linear momentum was removed to prevent center-of-mass drift. The system was equilibrated in the isothermal–isobaric (NPT) ensemble at a pressure of 1 bar, with temperature and pressure controlled using a Nosé–Hoover thermostat and barostat. A timestep of 2 fs was employed, and each simulation was run for 1,000,000 steps, corresponding to a total simulation time of \SI{2}{\nano\second}, which ensured sufficient sampling of the long-timescale interface dynamics.

This procedure was repeated across a series of target temperatures ranging from \SI{700}{\kelvin} to \SI{1420}{\kelvin}. To monitor the motion of the interface, the system was divided into two halves along the stacking direction, and trajectory snapshots were analyzed to determine the fraction of atoms exhibiting local face-centered cubic (FCC) ordering versus those classified as amorphous. This metric provides measures of the relative stability of the crystalline and disordered regions. At temperatures below the melting point, the crystalline fraction increased over time at the expense of the amorphous region, whereas at higher temperatures the amorphous fraction expanded. The melting temperature was identified as the point at which the interface remained stationary, with no net change in either density profiles or structural order parameters, thereby indicating equilibrium between crystalline and amorphous phases.

\section*{Acknowlegements}
This work was supported by the U.S. Department of Energy, Office of Science, Office of Basic Energy Sciences, Data, Artificial Intelligence, and Machine Learning at DOE Scientific User Facilities program under Award Number 34532 (Digital Twins). This work was performed in part at the Center for Nanoscale Materials and the Center for Nanophase Materials Sciences, which are U.S. Department of Energy, Office of Science User Facilities at Argonne National Laboratory and Oak Ridge National Laboratory, respectively. This work utilized the National Energy Research Scientific Computing Center, a DOE Office of Science User Facility supported by the Office of Science of the U.S. Department of Energy under Contract No. DE-AC02-05CH11231. We also acknowledge the LCRC computing facilities at Argonne.

\section*{Data Availability}
The codes, scripts, framework, and data that support the findings of this study are available from the authors upon reasonable request. The structure datasets used to train the SR1 and SR2 Potentials are available at https://github.com/bilvinv/Symbolic-copper .

\section*{Author Contribution}
 SKRS conceived the project. TDL and RB  developed the Symbolic regression framework with input from SKRS. BV deployed and extended the Symbolic regression workflow and trained the SR potential functions presented in this work. All the first-principles datasets were generated by SM and KB. Validation of the force fields was carried out by BV. SB and AK contributed to the calculation and analysis of the melting points and phonon dispersions presented in this manuscript. All the authors contributed to the data analysis and to the preparation of the manuscript. BV and SKRS wrote the manuscript with input from all the co-authors. BV, TDS, SB, AK, SM, KB, RB, MJC, OY, TP, BGS, and SKRS read and approved this manuscript. SKRS supervised and directed the overall project.

\section*{Competing Interests}
The authors declare no Competing Financial or Non-Financial Interests.

\newpage
\bibliography{ref_fixed}
\section*{Figure Legends}

\noindent\textbf{Figure 1.} The figure captures the training process for learning the Sutton-Chen Equation using EqNN a) The energy and force correlations showing very high correlations between the reference energy and force value calculated using SC EAM vs the ones predicted using EqNN b) The comparison between the pair term and the embedding term for the target equation and the one predicted by EqNN where for all relevant distances. c) The comparison of the Reference vs the predicted equations using the EqNN

\noindent\textbf{Figure 2.} Comparison of SC EAM predictions with DFT reference data for structures used to train the EqNN corresponding to the SC Equation. (a) Energy and (b) force predictions are shown across configurations ranging from near-ground state (left) to far-from-ground state (right). As the configurations deviate from equilibrium, both energy and force prediction errors (MAE) increase significantly.

\noindent\textbf{Figure 3.} Comparison of SR1 EAM predictions with DFT reference data for the dataset used to train the EqNN, using the same fixed embedding as the SC-EAM formalism. (a) Energy and (b) force predictions are shown across configurations ranging from near-ground state (left) to far-from-ground state (right). SR1 EAM exhibits significantly improved correlation with DFT energies and forces across all regimes compared to SC-EAM.

\noindent\textbf{Figure 4.} (a) Uniaxial equation of state (EOS) under compressive/tensile strain: SR1 (red) closely follows reference DFT (Ref; gray/black) across the full strain window, with a mean absolute error (MAE) of \(5.26\,\mathrm{meV/atom}\). (b) Volumetric EOS: SR1 maintains strong correlation with DFT (MAE \(=2.89\,\mathrm{meV/atom}\)). (c) Shear/tetragonal deformation: SR1 again agrees well with DFT (MAE \(=5.4\,\mathrm{meV/atom}\)).Insets in (a–c) depict the applied deformation modes. (d) Phonon dispersion along \(\Gamma\!-\!X\!-\!U\!-\!K\!-\!\Gamma\!-\!L\!-\!W\!-\!X\): SR1 reproduces the DFT branches with good agreement, particularly the acoustic slopes near \(\Gamma\); no imaginary modes are observed and a slight under-prediction of frequencies is noted. All energies are reported as \(\Delta E\) per atom referenced to unstrained fcc.

\noindent\textbf{Figure 5.} Comparison of SR2 EAM predictions with DFT reference data for the dataset used to train the EqNN, using the embedding function having more terms in addition to the square root term in the basis set. (a) Energy and (b) force predictions are shown across configurations ranging from near-ground state (left) to far-from-ground state (right). From left to right, we can see an increase in the range of energies where these structures fall, and we see a higher correlation in both energy and forces for all energy ranges compared with the predictions made by SR1.

\noindent\textbf{Figure 6.} (a–c) Change in energy per atom (\(\Delta E\)) versus strain for three deformation modes—(a) uniaxial, (b) volumetric (isotropic), and (c) shear showing SR2 (gold) closely tracking reference DFT (Ref; black) across \(\pm\,5 \%\) strain. Insets in (a–c) depict the applied deformation modes. (d) Phonon dispersion along \(\Gamma\!-\!X\!-\!U\!-\!K\!-\!\Gamma\!-\!L\!-\!W\!-\!X\): SR2 reproduces DFT with excellent agreement on the acoustic slopes near \(\Gamma\) and no imaginary modes, and overall shows improved correspondence relative to SR1. All energies are referenced to the unstrained fcc state.

\noindent\textbf{Figure 7.} The surface energy predictions. (a) Orientation-resolved surface energies \(\gamma(hkl)\) (bottom) for DFT (Ref) and the three force fields (SR2, SR1, SC), together with the corresponding absolute errors \(|\gamma_{\text{model}}-\gamma_{\text{DFT}}|\) (top). Across the set of low- and higher-index facets shown on the abscissa, SR2 and SR1 closely track the DFT values, while SC exhibits systematically larger deviations. Energies are in \(\mathrm{J\,m^{-2}}\). (b) Equilibrium Wulff constructions derived from the same \(\gamma(hkl)\) datasets. SR2 and SR1 reproduce the DFT polyhedron with similar facet types and relative areas, whereas SC shows noticeable changes in facet weighting. Facet colors correspond to the \([hkl]\) families indicated in the legend.

\noindent\textbf{Figure 8.} (a) Bain transformation path in Cu: energy per atom relative to fcc vs.\ $c/a$. All models give a minimum at fcc ($c/a=1$) and a maximum at bcc ($c/a=1/\sqrt{2}$); SR1/SR2 follow DFT closely, while SC overestimates the barrier. (b) Linear CTE, $\alpha \times 10^{6}$ (K$^{-1}$), vs.\ temperature: SR1/SR2 agree with SRM~736 within $\sim$5--10\% from 200–750~K, whereas SC is $\sim$40--70\% higher; curves are truncated at 750~K for comparability.

\noindent\textbf{Figure 9.} Comparison of model performance on polymorph properties. a) Polymorph energies relative to fcc: SR1/SR2 reproduce DFT ordering and magnitudes more accurately than SC, especially for higher-energy orthorhombic and trigonal phases. (b–c) Equation-of-state curves \((\Delta E\ \text{vs.}\ \text{strain})\) for BCC and HCP (eV/atom). SR1 and SR2 closely follow DFT near the minima and across the sampled strain range, while SC shows larger deviations, particularly for HCP. Insets illustrate the applied deformation.

\noindent\textbf{Figure 10.} Melting point calculations. (a) Initial two-phase simulation setup showing the solid-liquid interface and radial distribution functions (RDFs) for both phases. (b) Coexistence evolution curves for SR2 showing the relative amounts of solid and liquid after a \SI{1}{\nano\second} simulation at varying temperatures, with system snapshots at \SI{1200}{\kelvin}, \SI{1260}{\kelvin}, \SI{1300}{\kelvin}, and \SI{1400}{\kelvin}. The dashed black line marks the experimental melting point, and the dashed blue line indicates the SR2-predicted melting point. (c) Bar plot comparing melting point predictions from SR1, SR2, and Sutton-Chen EAM models. (d) Density evolution over time for simulations at \SI{1400}{\kelvin}, \SI{1300}{\kelvin}, and \SI{1200}{\kelvin}, showing phase transitions at each temperature.

\noindent\textbf{Figure 11.} Comparison of symbolic regression (SR)-derived potentials with the Sutton–Chen (SC) embedded-atom model (EAM). (a)Pair-wise interaction energies from SR2 (green), SR1 (red), and SC–EAM (blue) are plotted as a function of interatomic separation \(r\) (\si{\angstrom}). All three show the expected strong short-range repulsion and a single attractive well with \(V(r)\!\to\!0\) at large \(r\). SR1 and SR2 nearly coincide across the bonding region, while SC–EAM is stiffer at short range and exhibits a slightly deeper well depth near the minimum. (b) Embedding energy \(F_{\mathrm{emb}}(\rho)\) as a function of electron density \(\rho\) for SR2 (green) and SR1 (red). The two models coincide at low densities and progressively diverge with increasing \(\rho\), with SR2 yielding a more attractive (more negative) embedding energy at high \(\rho\). (c) Pointwise difference \(\Delta F_{\mathrm{emb}}(\rho)=F_{\mathrm{emb}}^{\mathrm{SR2}}(\rho)-F_{\mathrm{emb}}^{\mathrm{SR1}}(\rho)\), highlighting the density-dependent deviation (negative values indicate SR2 is deeper than SR1). Together, the panels show that SR2 introduces a stronger embedding response at higher local densities while preserving agreement in the low-\(\rho\) regime.

\noindent\textbf{Figure 12.}The Workflow for Symbolic regression. (a) Schematic of the nested sampling algorithm illustrating iterative exploration across the potential energy surface (PES). The vertical axis indicates potential energy, while the horizontal axis represents phase-space volume. Black dots mark the live set members distributed across phase space, with a dotted line highlighting the energy contour at each step. (b) Reduced-dimensional representation of the training-set latent space (SOAP descriptors), highlighting structural families and energy-landscape variations (c) Diagram of the EqNN neural network architecture, where each node corresponds to a distinct basis function that contributes to the overall model. (d) Visualization of the training process, showing iterative adjustments to optimize the network and improve model accuracy.
\bibliographystyle{unsrt}

\end{document}